\documentclass[fleqn,usenatbib]{mnras}

\usepackage{newtxtext,newtxmath}

\usepackage[T1]{fontenc}

\DeclareRobustCommand{\VAN}[3]{#2}
\let\VANthebibliography\thebibliography
\def\thebibliography{\DeclareRobustCommand{\VAN}[3]{##3}\VANthebibliography}

\usepackage{graphicx}	
\usepackage{amsmath}	
\usepackage{color,soul} 
\usepackage{verbatim}   
\usepackage{commath}

\usepackage{tikz,xcolor,hyperref}

\newcommand{\direct}{\texttt{DIReCT}}

\defcitealias{Chael_2018_ehtim}{C18}

\definecolor{lime}{HTML}{A6CE39}
\DeclareRobustCommand{\orcidicon}{%
    \begin{tikzpicture}
    \draw[lime, fill=lime] (0,0) 
    circle [radius=0.16] 
    node[white] {{\fontfamily{qag}\selectfont \tiny ID}};
    \draw[white, fill=white] (-0.0625,0.095) 
    circle [radius=0.007];
    \end{tikzpicture}
    \hspace{-2mm}
}

\newcommand{\orcidSamuel}{\href{https://orcid.org/0000-0001-9372-4611}{\orcidicon}}
\newcommand{\orcidNT}{\href{https://orcid.org/0000-0003-1602-7868}{\orcidicon}}

\title[Deep Learning Closure Imaging]{Deep Learning VLBI Image Reconstruction with Closure Invariants}

\author[S. Lai et al.]{Samuel Lai,$^{1}$\orcidSamuel\thanks{E-mail: samuel.lai@csiro.au}
Nithyanandan Thyagarajan,$^{1}$\orcidNT\,
O. Ivy Wong,$^{1,2}$\,
Foivos Diakogiannis,$^{1}$\,
and Lucas Hoefs$^{1,3}$\,
\\
$^{1}$Commonwealth Scientific and Industrial Research Organisation (CSIRO), Space \& Astronomy, P. O. Box 1130, Bentley, WA 6102, Australia\\
$^{2}$International Centre for Radio Astronomy Research, The University of Western Australia, Crawley, WA 6009, Australia\\
$^{3}$Mechatronics Engineering Department, Curtin University, Bentley, WA 6102, Australia\\
}

\date{Accepted XXX. Received YYY; in original form ZZZ}

\pubyear{2024}

\begin{document}
\label{firstpage}
\pagerange{\pageref{firstpage}--\pageref{lastpage}}
\maketitle

\begin{abstract}
Interferometric closure invariants, constructed from triangular loops of mixed Fourier components, capture calibration-independent information on source morphology. While a complete set of closure invariants is directly obtainable from measured visibilities, the inverse transformation from closure invariants to the source intensity distribution is not established. In this work, we demonstrate a deep learning approach, Deep learning Image Reconstruction with Closure Terms (\direct), to directly reconstruct the image from closure invariants.
Trained on both well-defined mathematical shapes (two-dimensional gaussians, disks, ellipses, $m$-rings) and natural images (CIFAR-10), the results from our specially designed model are insensitive to station-based corruptions and thermal noise. The median fidelity score between the reconstruction and the blurred ground truth achieved is $\gtrsim 0.9$ even for untrained morphologies, where a unit score denotes  perfect reconstruction. In our validation tests, \direct's results are comparable to other state-of-the-art deconvolution and regularised maximum-likelihood image reconstruction algorithms, with the advantage that \direct\ does not require hand-tuned hyperparameters for each individual prediction. This independent approach shows promising results and offers a calibration-independent constraint on source morphology, ultimately complementing and improving the reliability of sparse VLBI imaging results. 
\end{abstract}




\section{Introduction} \label{sec:Introduction}

Radio interferometry is a technique that utilises an array of receiver elements to measure spatial correlations of incident radiation. Through image synthesis in radio interferometry, these measured correlations are leveraged to infer the intensity distribution of sources on the sky plane. Essentially, the radio interferometer emulates a single-dish telescope with an aperture equivalent to the length of its longest baseline. Consequently, the longest baseline dictates the smallest resolution element probed by the array, thereby defining the nominal image resolution. The smallest of scales are therefore probed by arrays separated on continental or planetary scales, characteristic of Very Long Baseline Interferometry (VLBI). However, image reconstruction from VLBI data can be extremely challenging due to the sparsity of measurements, necessitating accurate calibration of signals obtained from heterogeneous array elements dispersed over considerable distances \citep{TMS}. 

Recently, the Event Horizon Telescope Collaboration \citep[EHTC;][]{Doeleman_2009} utilised VLBI measurements on Earth-sized baselines to image the black hole event horizons of the central supermassive black holes within M87 \citep{EHT_2019_Imaging} and Sgr A* \citep{EHT_2022_SgrAImaging}. The collaboration followed a detailed verification process to ensure that the result is robust against biases present in parameter or algorithmic selection \citep[e.g.][]{Shepherd_2011_DIFMAP, Akiyama_2017_imaging, Chael_2018_ehtim} and is recovered under a broad range of imaging parameters. Subsequent analyses by independent groups have also found consistent results \citep[e.g.][]{Arras_2022, Broderick_2022, Carilli_2022, Lockhart_2022, Muller_2024_closureTraces, Feng_2024}. However, finer details of the reconstruction, including the ring thickness and surface brightness dynamic range, can vary substantially between different methodologies or different imaging assumptions \citep[e.g.][]{EHT_2019_Imaging}. \citet{Carilli_2022} showed that while an image of a ring with higher surface brightness in the southern half leads to the lowest noise and residuals, structures that are not clearly ring-like can be produced with different choices of the initial model. This underscores how differences in the calibration process can lead to divergent results, necessitating an accurately calibrated set of measurements, which is itself a finely-tuned iterative process.

It is possible to bypass major elements of the calibration process with specific arrangements of the interferometric measurements, such as closure phases \citep{Jennison_1958} or closure amplitudes \citep{Twiss_1960}. These ``closure quantities'' are immune to corruption from individual station-based noise properties, derived from  propagation delays and clock errors introducing phase variations, or pointing errors and flux calibration introducing gain errors. Closure quantities serve as calibration-independent true observables, which carry robust information about the source properties only limited by thermal noise and non-station-based errors. Aside from their advantages in co-polar interferometry, closure quantities also carry information on a source's intrinsic polarisation \citep{Broderick_2020, Samuel_2022}, which are independent of leakage between polarisation feeds. Developments of regularised maximum-likelihood methods \citep[e.g.][]{Ikeda_2016, Akiyama_2017_polarimetric, Akiyama_2017_imaging, Chael_2018_ehtim, Blackburn_2020}, building off of the earlier maximum entropy method \citep[e.g.][]{Frieden_1972, Narayan_1986}, have shown that closure quantities can be leveraged directly for imaging by minimising an objective function containing closure data terms. Similarly, closure quantities have been used as a constraint in compressive sensing techniques which encode the image with a sparse set of basis functions representing various spatial scales \citep[e.g.][]{Mertens_2015, Mueller_2022_doghit, Muller_2024_closureTraces}. 

The direct mapping from closure quantities to the object’s morphology is therefore of considerable interest. Previous work utilised machine learning to capture the intricate non-linear relationships between closure quantities and the source intensity distribution, allowing the trained networks to classify image morphology \citep{Thyagarajan_2024_Lucas} or directly reconstruct the image given a prior to impose desired image characteristics \citep{Feng_2024}. Normalising flow methods paired with generative networks have also been shown to be useful in characterising multi-modalities in VLBI image reconstructions, enabling more efficient posterior estimation \citep{Sun_2020, Sun_2022}. In this paper, we present, for the first time, that the co-polar closure invariants formalism described in \citet{Thyagarajan_2022_CI} and exploited in \citet{Thyagarajan_2024_Lucas} for morphological classification, can be leveraged to perform direct image reconstruction with a supervised machine learning approach, which we call \direct, for Deep learning Image Reconstruction with Closure Terms. We use a transformer to extract features from interrelationships within the set of closure invariants with global contextual awareness. Then, we validate our model and quantify its performance under different conditions with noise models consisting of both station-based corruptions and thermal noise. Moreover, we consider how our model trained on predominantly natural images performs for general untrained astrophysically-relevant morphologies. Finally, we compare our reconstructions to state-of-the-art algorithms.

The content of this paper is organised as follows: in Section \ref{sec:vis-closure}, we introduce the co-polar radio interferometry context and closure invariant formalism. In Section \ref{sec:imrec}, we describe established image reconstruction pipelines and present our machine learning architecture, which performs the image reconstruction task with closure invariants. We also describe our training dataset, loss functions, and training strategy. In Sections \ref{sec:results} and \ref{sec:discussion}, we present and discuss the image reconstruction results alongside products from state-of-the-art algorithms, which we show for comparison. We present a summary of this work and conclusion in Section \ref{sec:conclusion}. 

\section{Visibilities and Closure Invariants} \label{sec:vis-closure}

Radio interferometers measure the coherence of received signals from distant sources, the complex visibility function $\mathcal{V}(u,v)$, which is related to the source intensity distribution, $T(l,m)$, by the 2-D Fourier transform, $\mathcal{F}$, under reasonable assumptions \citep{TMS},
\begin{equation}
    \mathcal{V}(u,v) = \mathcal{F}\{T(l,m)\} = \iint T(l,m) e^{-i2\pi (ul + vm)} dl dm\,,
\end{equation}
where, $(u,v)$ are spatial frequencies measured in wavelengths and $(l, m)$ are angular coordinates expressed in direction cosines corresponding to the projection of the celestial sphere onto the tangent sky plane. Each complex visibility, $\mathcal{V}(u,v)$, expressed as an amplitude and a phase, is therefore a measurement of a single Fourier component of the source intensity distribution, $T(l,m)$, in the $(l,m)$ domain. 
One baseline formed between a pair of stations measures a single instantaneous visibility, and the total number of independent visibilities from $N_s$ stations is given by the binomial coefficient,
\begin{equation}
    N_{\rm{vis}} = \binom{N_s}{2} = \frac{N_s(N_s - 1)}{2}\,.
\end{equation}

Consider the case in which each array element's measurement is corrupted by a time-dependent multiplicative complex factor, $g_a$, representing both amplitude and phase distortions such that the measured visibility between array elements, $(a,b)$, is given by $\mathcal{V}'_{ab} = g_a \mathcal{V}_{ab} g_b^* + \epsilon_{ab}$, where $\epsilon_{ab}$ is a zero-mean thermal noise term. Although such corruptions are represented here as station-based effects, in practice, measured visibilities can be further biased by polarisation leakage, cross-talk, and electronic imperfections in the feed, among other effects. For the purposes of this paper, we consider only station-based systematic errors and co-polar interferometry.

\subsection{Closure Invariant Formalism}
The construction of special ``closure quantities'', which are insensitive to arbitrarily large station-based corruptions,  have been employed in radio astronomy for decades \citep[e.g.][]{Jennison_1958, Twiss_1960}. The set of independent closure quantities, composed of ``closure phases'' and ``closure amplitudes'', are created from different combinations of Fourier components, which can be difficult to interpret physically and provide less overall information on the source than the full set of calibrated complex visibilities. Nevertheless, closure quantities have been integral to advances in calibration and synthesis imaging in interferometric radio astronomy \citep[e.g][]{Rogers_1974, Readhead_1978, Cornwell_1999}. 

Closure phases are constructed from a triangle of baselines which eliminates the complex gain phase terms. By choosing a reference triangle vertex, one can construct $N_{\Delta} = (N_s-1)(N_s-2)/2$ independent closure phases \citep{TMS}, defined for stations $(a,b,c)$ as
\begin{equation} \label{eq:closure_phases}
    \mathcal{B}'_{\Delta} = \mathcal{V}'_{ab}\mathcal{V}'_{bc}\mathcal{V}'_{ca}\,,
\end{equation}
which is a complex quantity called the visibility bispectrum. The phase term of the bispectrum (also known as closure phase) is preserved under any station-based phase errors, and apart from additive thermal noise, the measured closure phase is a robust measurement of the true closure phase of the observed image. 

Similarly, one can construct up to three closure amplitude terms for a quadrilateral loop of four array elements, $(a,b,c,d)$, 
\begin{align} 
\begin{split} \label{eq:closure_amps}
    \mathcal{R}'_{\square,1} = \frac{\abs{\mathcal{V}'_{ab}}\abs{\mathcal{V}'_{cd}}}{\abs{\mathcal{V}'_{ac}}\abs{\mathcal{V}'_{bd}}}\,,
    \mathcal{R}'_{\square,2} = \frac{\abs{\mathcal{V}'_{ac}}\abs{\mathcal{V}'_{bd}}}{\abs{\mathcal{V}'_{ad}}\abs{\mathcal{V}'_{bc}}}\,,
    \mathcal{R}'_{\square,3} = \frac{\abs{\mathcal{V}'_{ad}}\abs{\mathcal{V}'_{bc}}}{\abs{\mathcal{V}'_{ab}}\abs{\mathcal{V}'_{cd}}}\,,
\end{split}
\end{align}
only two of which are independent (note that $\mathcal{R}'_{\square,1}\mathcal{R}'_{\square,2}\mathcal{R}'_{\square,3}=1$), thereby producing a total of $N_{\square} = N_s(N_s-3)/2$ independent closure amplitudes \citep{TMS,Cornwell_1999}. The amplitude of the quadrilateral construction is robust against amplitudes of the gain terms, but it is likewise affected by additive noise in the visibilities. 

For $N_s$ stations, the total number of independent closure phases and amplitudes is always fewer than independent visibility phases and amplitudes by a factor of $\left(1-\frac{2N_s-1}{2N_{\rm{vis}}}\right)$, which results in a loss of information. Notably, closure phases and amplitudes do not preserve information on the source's absolute position or total flux density. As such, these parameters need to be constrained through separate measurements or with biasing decisions in the image reconstruction (\textit{i.e.} by centering the image and rescaling the flux). 

Recently, \citet{Thyagarajan_2022_CI} and \citet{Samuel_2022} developed a general unified formalism of interferometric closure invariants for co-polar and polarimetric measurements, respectively. In this study, we focus on the former Abelian gauge theory formalism to obtain a complete and independent set of closure invariants. Briefly, the closure invariant formalism identifies $N_{\Delta}$ triangular loops, all pinned on a reference vertex indexed as 0, and defines an \textit{advariant} associated with each triangle as, 
\begin{equation}
    \mathcal{A}'_{\rm{0ab}} = \mathcal{V}'_{0a}(\mathcal{V}'^{*}_{ab})^{-1}\mathcal{V}'_{b0} = \abs{g_0}^2\mathcal{A}_{\rm{0ab}}\,,
\end{equation}
where the unknown scaling factor on all complex advariants, $\abs{g_0}^2$ associated with the reference vertex, can be canceled by normalising to any one non-zero advariant quantity, resulting in the loss of one degree of freedom and a total of $N_{\Delta} + N_{\square} = N_s^2 - 3N_s + 1$ independent real-valued closure invariant quantities. The traditional $N_{\Delta}$ closure phases and $N_{\square}$ amplitudes in Equations \ref{eq:closure_phases} and \ref{eq:closure_amps} can be identified directly from pairs of complex advariants. Further details, including advantages of this formalism, are presented in \citet{Thyagarajan_2022_CI} and the polarimetric extension is presented in \citet{Samuel_2022}.

\section{Sparse VLBI Image Reconstruction} \label{sec:imrec}
In this section, we describe image reconstruction pipelines with state-of-the-art algorithms: standard \texttt{CLEAN} \citep{Hogbom_1974_CLEAN}, \texttt{eht-imaging} \citep{Chael_2018_ehtim}, and \texttt{DoG-HiT} \citep{Mueller_2022_doghit}. We then present our deep learning architecture which performs the image reconstruction task using the set of closure invariants derived from the \citet{Thyagarajan_2022_CI} formalism, an approach that we call \direct, for Deep learning Image Reconstruction with Closure Terms. We describe our image dataset, customised loss functions, and training strategy to produce our final model for comparison to state-of-the-art pipelines in Section \ref{sec:discussion}. 

\begin{figure*}
	\includegraphics[width=0.9\textwidth]{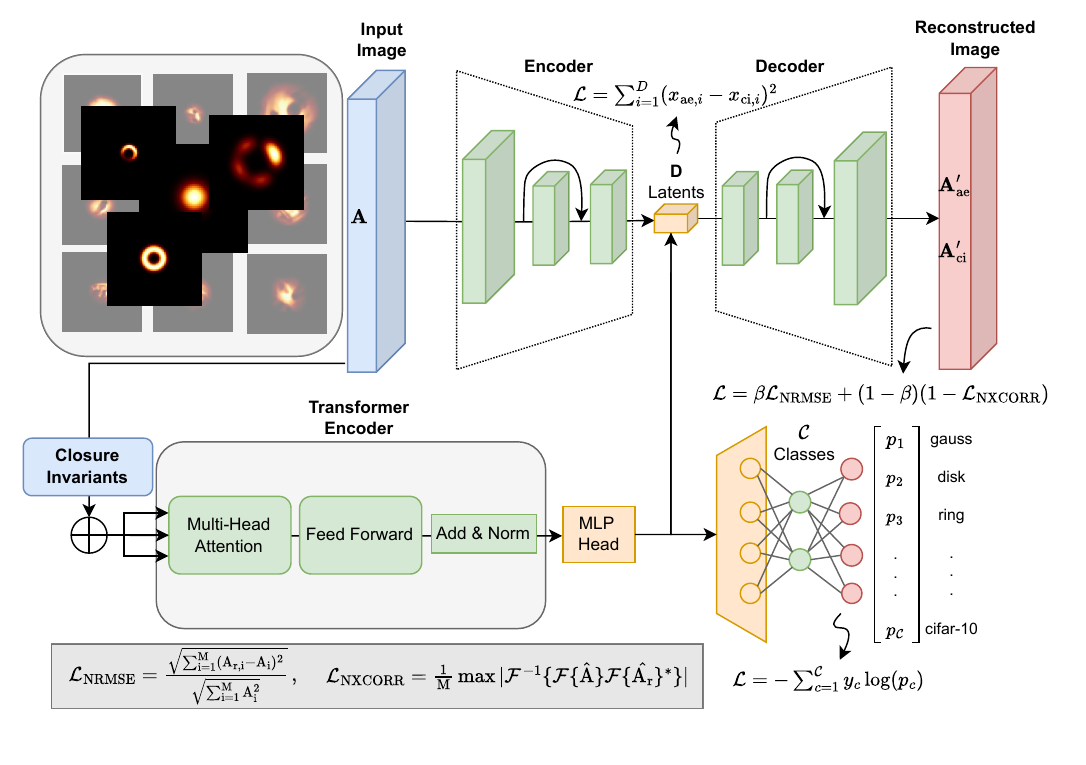}
    \caption[Machine learning architecture]{Illustration of the machine-learning architecture adopted in this study. During training, images are fed into a convolutional encoder with residual connections to produce a compressed information-rich representation, which is also predicted by a transformer encoder that accepts a set of closure invariants produced by synthetic observation with a VLBI array as input. The latent space can be used to reconstruct the image using the convolutional decoder and predict the morphological class with the multilayer perceptron. Once trained, image reconstructions can be produced directly from a set of closure invariants using the transformer and decoder. We describe the loss functions associated to each output with $\mathcal{L}$.}
    \label{fig:ML-architecture}
\end{figure*}

\subsection{CLEAN}
\texttt{CLEAN} is one of the most successful and influential deconvolution algorithms. Originally devised by \citet{Hogbom_1974_CLEAN}, multiple modern variations of \texttt{CLEAN}  have been proposed since and continue to be extensively used in radio astronomy \citep[e.g.][]{Wakker_1988, Cornwell_2008_CLEAN, Offringa_2014_WSCLEAN}. \texttt{CLEAN}  iteratively models the brightest points in the dirty image as point sources, subtracting a scaled response from the data until no significant sources remain. The final reconstructed image is the cleaned model convolved with a clean restoring beam alongside the final residual map. Although \texttt{CLEAN} is a highly non-linear algorithm, \citet{Schwarz_1978} showed that under specific conditions, \texttt{CLEAN} converges to a least-mean-squares fit of the Fourier transformed delta function components to the measured visibility. Later, \citet{Marsh_1987} showed that when applied to intensity images, \texttt{CLEAN} can be regarded as an approximate method that minimises the total flux consistent with observations, where the method becomes exact when the data consists of a point source.

The final image reconstruction from \texttt{CLEAN} is affected by various subjective decisions, such as the gain used to adjust the dirty beam response, the window area selected for image processing, and the termination criteria \citep[e.g.][]{Taylor_SIRAII_1999}. Moreover, in instances of uncertain calibration,  the conventional ``self-calibration'' approach introduces additional adjustment mechanisms, including the initial source model employed for calibration, the complex error model, and frequency of recalibration. Each of these finely tuned systems can influence the convergence of the final product. Notably in the context of VLBI imaging with the EHT, self-calibration applied to the highly unstable phases biases the procedure towards the prior model, potentially resulting in markedly different output morphologies \citep[e.g.][]{Carilli_2022} due to the multimodality of the posterior landscape \citep[e.g.][]{Muller_2023_Multiobj}. This further highlights how closure-only imaging, which are less affected by such corruptions, is valuable in VLBI imaging. Throughout Section \ref{sec:discussion}, we apply \texttt{CLEAN} on noiseless synthetic observations with a 0.1 gain on the central half of the image, and run for a fixed 1000 iterations. As our synthetic data is noiseless, we do not require a prior model for self-calibration.

\subsection{eht-imaging}

The \texttt{eht-imaging} algorithm \citep{Chael_2018_ehtim} is in a class of regularised maximum-likelihood forward-modelling methods \citep[e.g.][]{Ikeda_2016, Akiyama_2017_polarimetric, Akiyama_2017_imaging}, building off from the more familiar and traditional Maximum Entropy Method \citep[MEM;][]{Frieden_1972, Gull_1978, Cornwell_1985, Narayan_1986}. Regularised maximum-likelihood methods operate by minimising an objective function composed of weighted data terms and regularisers. By minimising the $\chi^2$ goodness-of-fit in the visibilities, bispectrum, or closure quantities, only a single forward Fourier transform is required to transform model images to the visibility domain. In \texttt{eht-imaging}, multiple data terms and regularisers can be minimised simultaneously, similar to optical interferometry techniques \citep[e.g.][]{Baron_2010SPIE}. Regularisers implemented in \texttt{eht-imaging} include entropy \citep{Frieden_1972, Gull_1978, Narayan_1986}, the sparsity-promoting $l1$-norm \citep{Honma_2014}, isotropic total variation \citep{RUDIN1992259}, total squared variation \citep{Kuramochi_2018}, total image flux density, and image centroid position. Unlike the procedural methods, such as \texttt{CLEAN} and its variants, regularised maximum-likelihood deconvolution algorithms require setting an image prior \citep{TMS} which can affect the resulting image reconstruction in some circumstances \citep[e.g.][]{Carilli_2022}. While \texttt{CLEAN} implicitly assumes that the source can be approximated as a collection of point sources, regularized maximum likelihood methods such as \texttt{eht-imaging} typically promote smoothly varying surface brightness distributions through the use of regularization terms. However, the actual behavior of these methods can vary depending on the specific implementation and chosen regularization parameters.

In Section \ref{sec:discussion}, we initialise all regularised maximum-likelihood image reconstruction pipelines with a 40\,$\mu$as full-width at half maximum (FWHM) centralised Gaussian prior and a five magnitude fainter Gaussian with the same FWHM, offset by 40\,$\mu$as in each axis, to break the symmetry and avoid gradient singularities (refer to EHT M87* 2019 Imaging Pipeline\footnote{\href{https://github.com/eventhorizontelescope/2019-D01-02}{https://github.com/eventhorizontelescope/2019-D01-02}}). The specific implementation of the \texttt{eht-imaging} pipeline with visibilities includes four imaging rounds, which differ only in their regularisation terms. The first and last imaging rounds utilise the entropy regulariser and the central two rounds use the isotropic total variation regulariser. All imaging rounds also include the total image flux density and centroid position regularisers, but the weighting of the flux density regulariser is double that of the centroid regulariser. Between imaging rounds, the intermediate reconstructed image is blurred by a circular beam with a radius of the nominal array resolution to help with convergence and smooth over spurious super-resolved features. After the last imaging round, the final product is blurred with half of the effective beam, which is intended to blur out spurious high-frequency structure.

\subsubsection{Closure-only imaging}
Closure-only image reconstructions can be produced from \texttt{eht-imaging} by limiting the data terms in the objective function to (log)-closure amplitudes and closure phases, where log closure amplitudes are identified as more robust observables for imaging applications \citep{Chael_2018_ehtim}. We implement the closure-only image reconstruction pipeline in Section \ref{sec:discussion} with five imaging rounds. To aid in convergence, the first three imaging rounds includes down-weighted corrupted visibilities in the initial minimisation steps and we employ equally weighted closure phase and log closure amplitude data terms in all imaging iterations. The sole deviation from this strategy occurs during the initial imaging round, wherein closure phases are assigned a weight twice that of the log closure amplitudes as they are more valuable for producing sensible image priors for subsequent imaging rounds. All imaging rounds use the total squared variation regulariser, except the first round which uses simple entropy. 

Each imaging round also incorporates regularization terms for total image flux density and image centroid. Intermediate images undergo convolution with a circular beam, followed by a final convolution with half the effective beam after the final imaging round. Corrupted visibilities play a minor role in the initial convergence and have minimal impact on final product of the closure-only imaging pipeline, rendering the output robust against multiplicative noise such as station-based gains. While convolving the final product from a regularised maximum likelihood approach with the beam is not a principled approach, we find that it produces the best quantitative and qualitative results from the \texttt{eht-imaging} pipeline in both reconstructions based on visibilities and closure quantities. An alternative is optimising the relative weighting of the smoothness regulariser, which would also suppress the emergence of high-frequency artifacts. However in this study, we choose not to independently fine-tune the image reconstruction pipeline for each individual reconstruction to avoid conferring any one algorithm an unfair advantage during comparison.

\subsection{DoG-HiT} \label{sec:doghit}
\begin{figure*}
	\includegraphics[width=0.95\textwidth]{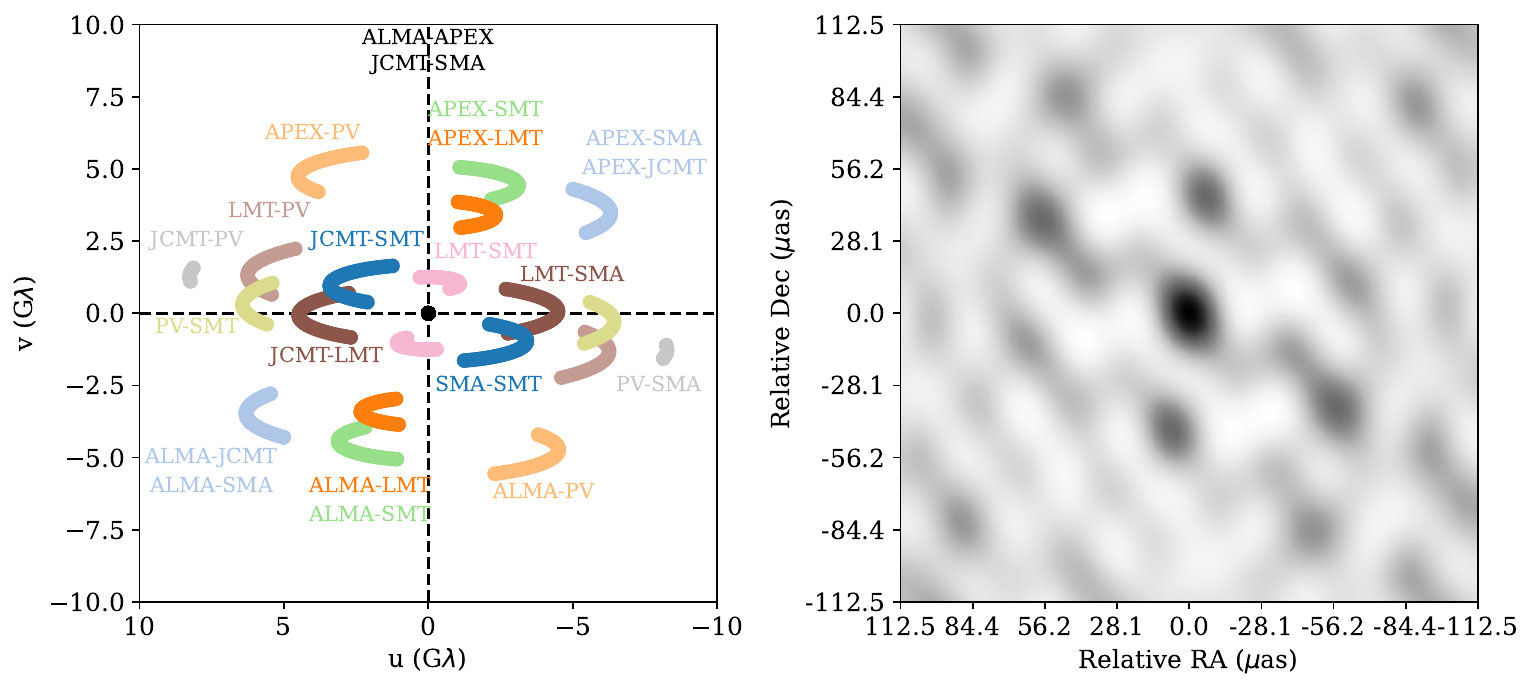}
    \caption[Sampling and dirty beam]{Illustration of the sampling function on the $uv$-plane under rotation synthesis for a pointing location of $\rm{RA = 12^h30^m49.42^s\, and \, Dec = +12^{\circ}23'28.04''\, (J2000)}$, and the resulting dirty beam. Observations are synthesised with short 5s integrations every 10 minutes over a period of 24 hours on 2017-04-05. Rotation synthesis curves are labeled with the associated station pairs. All images shown in this study share the same angular scale as the dirty beam image ($225\times 225\,\mu$as).}
    \label{fig:uv-sampling}
\end{figure*}

\texttt{DoG-HiT} \citep[Difference of Gaussian $-$ Hard image Thresholding;][]{Mueller_2022_doghit} is a multiscalar wavelet regularised maximum-likelihood algorithm built on compressive sensing techniques \citep[e.g.][]{Wiaux_2009, Li_2011_CompSens, Garsden_2015, Pratley_2018}. By utilising a flexible data-driven dictionary of multiscalar difference-of-Gaussian wavelet functions, \texttt{DoG-HiT} achieves super-resolution in comparison to CLEAN and has the potential to match or outperform regularised maximum-likelihood methods, while requiring fewer parameters \citep{Mueller_2022_doghit}. The key assumption is that the source structure can be sparsely represented by a few wavelets.  

We briefly summarise the five separate imaging rounds in the \texttt{DoG-HiT} imaging pipeline applied for image reconstruction in Section \ref{sec:discussion}. The first imaging round uses \texttt{eht-imaging} to simultaneously minimise the visibility amplitude, closure phase, and closure amplitude residuals with total scalar flux, image centroid, and entropy regularisation terms. The result from the first imaging round, convolved with the effective beam, produces a sensible initial guess of the source intensity distribution from the Gaussian prior and reduces overall computation time in subsequent imaging rounds. The second imaging round optimises the wavelet coefficient array to minimise closure phase and amplitude residuals with scalar flux and hard threshold sparsity regularisation terms. The largest scales are considered first and smaller scales are added successively with larger thresholds. After self-calibrating with the result from the second imaging round, the third imaging round includes the visibility amplitudes as a data term alongside the closure quantities with a ``multi-resolution support'' regularisation term, which acts as a compact flux constraint. The fourth and final imaging rounds include the complex visibilities and thresholds the image pixels to only positive values to refine the final image reconstruction. We make minimal adjustments to the imaging pipeline and parameters described in \citet{Mueller_2022_doghit} in order to adapt to our $uv$-coverage.

\subsection{\direct\ - Deep Learning with Closure Invariants} \label{sec:conv_ae}

\begin{figure*}
	\includegraphics[width=0.9\textwidth]{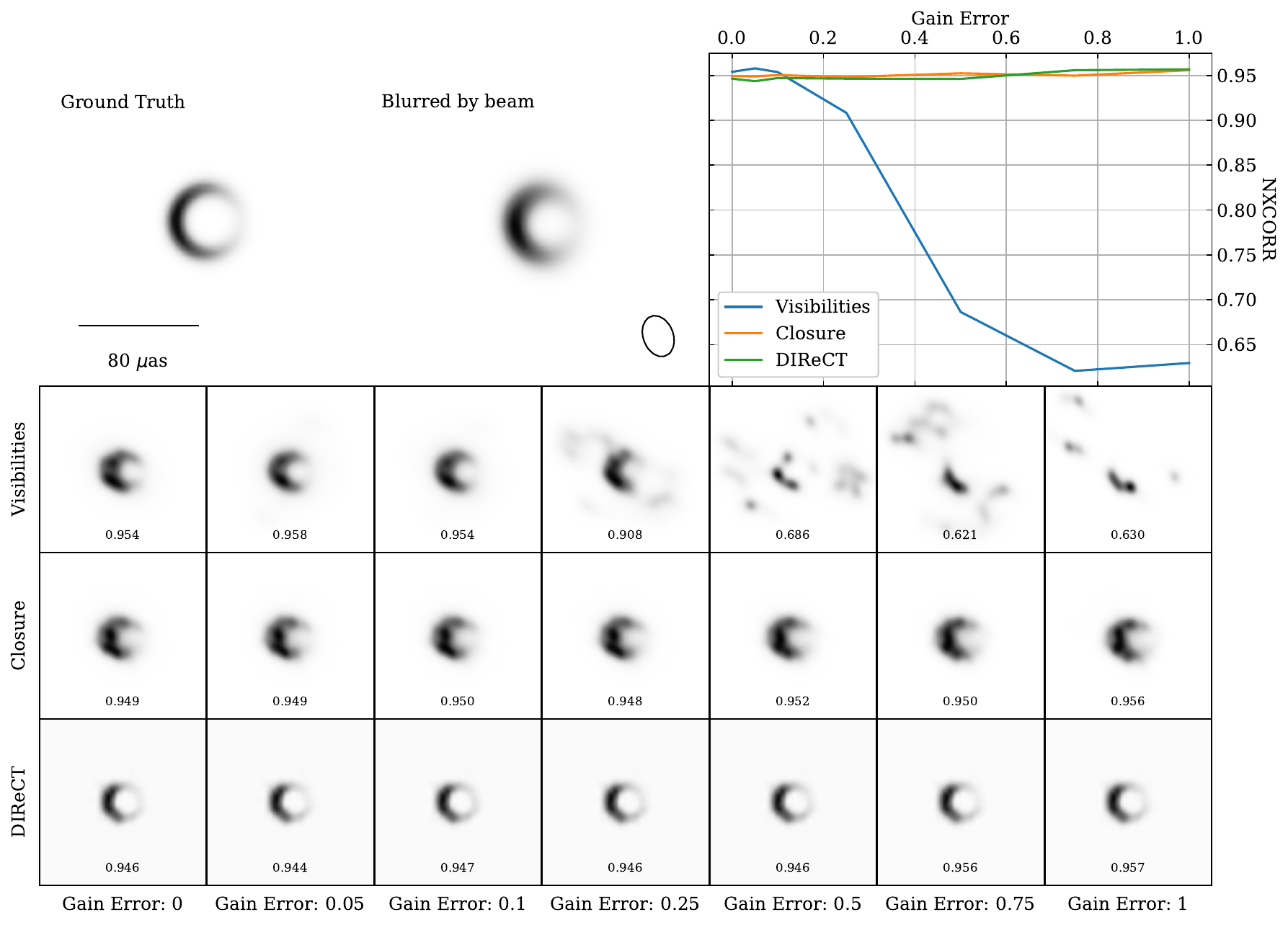}
    \caption[Effect of gain error]{(Top left) Ground truth image of a crescent represented by a first order $m$-ring \citep{Roelofs_2023_mring}. (Top middle) Ground truth blurred by the effective beam. (Top right) Illustration of the sensitivity in the normalised cross-correlation fidelity to gain error for regularised maximum-likelihood reconstructions using visibilities and closure invariants using \texttt{eht-imaging}, compared with our deep learning model (\direct). (Middle rows) Image reconstructions of each method for different levels of amplitude gain error. NXCORR fidelity scores are shown underneath each reconstructed image. All reconstructions are affected by random, zero-mean thermal noise.}
    \label{fig:gain-noise}
\end{figure*}

Although the transformation from an image to a complete independent set of closure invariants for any interferometric array is deterministic, the inverse transformation is not, resulting in an ill-posed problem with potentially infinite viable solutions due to the sparsity in the $uv$-coverage. We confront this problem with a deep learning approach that performs the mapping from closure invariants back to images, calling the approach: \direct, for Deep learning Image Reconstruction with Closure Terms.

Our machine learning architecture, shown in Figure \ref{fig:ML-architecture}, is principally composed of a convolutional autoencoder \citep[e.g.][]{KRAMER1992313} with skip connections. A convolutional autoencoder is composed of layers of convolutional and pooling operations, followed by upsampling and deconvolution layers for decoding. Skip connections are incorporated to bypass certain layers within the neural network architecture by enabling the direct transmission of information from earlier layers to later ones, which helps to alleviate the degradation problem commonly encountered during deep learning training, where an increase in the depth of a network leads to a decrease in performance. The degradation problem is caused by the difficulty in optimising a large numbers of parameters in a deep network, but skip connections provide the identity mapping of the outputs between layers, emulating a shallower network if that is optimal for the problem. This network encodes an input image into a compressed information-rich representation, which is then fed into a transposed convolutional decoder to reconstruct the image. A morphological class label from one of the input classes is predicted by the latent features through a feedforward multilayer perceptron, composed of layers of interconnected neurons. 

Noiseless synthetic observations of the input image are produced with a sparse VLBI array and deterministically reduced into a complete independent set of closure invariants with the formalism of \citet{Thyagarajan_2022_CI}. The closure invariants are fed into a transformer encoder \citep{Vaswani_2017} which enables the model to learn features from interrelationships within the data with global contextual awareness. The transformer encoder consists of positional encoding of the input, four layers of multi-head attention modules, and feedforward networks which facilitate extraction of meaningful features from the input data for downstream tasks. By passing the output through a final multilayer perceptron head, the transformer directly predicts the latent features, which can be decoded to reconstruct the image and predict the morphological class. Once trained, the model produces image reconstructions from a set of closure invariants through the transformer and decoder. Furthermore, the class label can be simultaneously predicted from the latent features through the trained multilayer perceptron classifier. Note that the pipeline for prediction tasks never accesses the image data, visibilities, or $uv$-coverage.  

In the purpose-designed model, the convolutional autoencoder guarantees that the presence of a compressed representation in the latent features for the transformer encoder to forecast for any input morphology, even those not represented in the training dataset, improving the model's generalisability and adaptability. Furthermore, by including the class prediction multi-layer perceptron, certain sets of properties in both the image space and closure invariants can be associated with a particular class, thereby facilitating improved image reconstruction when a high-confidence class label can be assigned. However, removing the classifier does not significantly influence the performance of the image reconstruction task and we do not place significant emphasis on the class prediction when reconstructing images from closure invariants, particularly for untrained morphologies that can not be definitively assigned a class. The four components of the loss function correspond to the class prediction cross-entropy loss, mean squared error loss between two sets of predicted latent quantities, and two image reconstruction losses. We describe the loss functions in greater detail in Section \ref{sec:training_network} where we discuss training the network.

\subsubsection{Training dataset}
The training dataset consists of extended sources, such as Gaussians, disks, rings, ellipses, and $m$-rings of first and second order \citep{Roelofs_2023_mring}, each convolved with a Gaussian blurring filter with random FWHM from $10-50\,\mu$as. Disks differ from Gaussian morphologies in that they are initially defined with a unit intensity within a selected radius and zero everywhere else.  Our dataset also includes multiple sources, such as double disks and double Gaussians. Though the training dataset includes $\sim 1000$ unique images from each of the aforementioned distinct morphological classes, the vast majority of the dataset is composed of CIFAR-10 \citep{Krizhevsky09_CIFAR10}, which is a set of 60,000 images representing ten natural object classes (e.g. birds, motor vehicles, cats). In \citet{Feng_2024}, CIFAR-10 was found to be the least biased training dataset for VLBI image reconstruction algorithms when compared to images derived from general relativistic magnetohydrodynamic (GRMHD) simulations or the CelebA collection of celebrity faces \citep{Liu_2015_CelebA}. The combined dataset is therefore composed of 67,404 images, 20\% of which are set apart for validation. The CIFAR-10 dataset is considered as a single morphological class in our network. 

To further expand the variety of images encountered by the model, we introduce random augmentations of all images during training. The augmentations encompass vertical or horizontal flips, rotation, rescaling down to $80\%$ of the original source size, and a $5^{\circ}$ shear. To preclude padding the image where no data is available which introduces flux discontinuities, our original images are $128\times128$ pixels, which are then cropped to $64\times64$ pixels subsequent to augmentation. We use bilinear interpolation for non-discrete shifts and rotations to maintain smoothness in the augmented dataset. The CIFAR-10 images include an additional radial tapering augmentation in order to incorporate the assumption of a centralised object with a black background. The taper is defined as an element-wise multiplication between the original image with a mask composed of a disk blurred by a Gaussian kernel. The mask disk radius is randomised between $28-84\,\mu$as and and the Gaussian kernel FWHM is randomised between $18-35\,\mu$as.

\subsubsection{Training the network} \label{sec:training_network}

The training of our network is guided by a combination of four loss terms weighted by an associated hyperparameter vector, $\vec{\alpha}$. Each loss term corresponds to an output from the network: a class prediction ($\mathcal{L}_{\mathcal{C}}$), two independent sets of latent variables ($\mathcal{L}_{D}$), and two independent reconstructed images for each input image ($\mathcal{L}_{\rm img}$). The full loss function for an image \textbf{A} and class label $c$ is described by,
\begin{equation}
   \mathcal{L}{\left({\rm \textbf{A}}, c\right)} = \vec{\alpha} \cdot \begin{bmatrix}
             -\sum_{c=1}^\mathcal{C} y_{c}\log(p_{c})\\
             \sum_{i=1}^{D}(x_{{\rm ae},i}-x_{{\rm ci},i})^2 \\ 
             \mathcal{L}_{\rm img}({\rm \textbf{A}, \textbf{A}_{ae}'})\\
            \mathcal{L}_{\rm img}({\rm \textbf{A}, \textbf{A}_{ci}'})
         \end{bmatrix} \,, \label{eq:losses}
\end{equation}
composed of the cross-entropy loss for $\mathcal{C}$ classes, the mean squared error loss between two sets of $D$ predicted latent quantities, and two image reconstruction losses. Cross-entropy and mean squared error loss functions are proper scoring metrics for maximum likelihood estimation, for Bernoulli and Gaussian distributed variables, respectively. With respect to the image reconstruction loss, one common technique is to extend the mean square error concept to two dimensions and compute the simple normalised root-mean-square error (NRMSE) fidelity metric, following \citet{Chael_2016}, \citet{Akiyama_2017_imaging}, and \citet{Chael_2018_ehtim}. The NRMSE metric evaluates two images \textbf{A} and \textbf{B} on their pixel-by-pixel similarity,
\begin{equation}
    {\rm NRMSE(\textbf{A}, \textbf{B})} = \frac{\sqrt{\sum_{i=1}^{M} (A_i - B_i)^2}}{\sqrt{\sum_{i=1}^{M} B_i^2}}\,,
\end{equation}
where, image \textbf{A} is evaluated with respect to \textbf{B} for $M$ pixels each. However, closure-only image reconstruction methods are insensitive to the true position of the source centroid within the field of view. Accurate closure-only reconstructions can still score poorly compared to methods which can optimise the calibrated visibility phases. Therefore, we also consider the normalised cross-correlation (NXCORR) fidelity metric \citep{EHT_2019_Imaging},
\begin{equation}
    {\rm NXCORR(\textbf{A}, \textbf{B})} = \frac{1}{M}\max\abs{{\mathcal{F}^{-1}\{\mathcal{F}\{\hat{A}\}\mathcal{F}\{\hat{B}\}^*\}}}\,,
\end{equation}
where $\mathcal{F}$ and $\mathcal{F}^{-1}$ are forward and inverse Fourier transforms, respectively, and $\hat{A} = (A - \bar{A})/\sigma_A$ is the image $A$ normalised by the mean and standard deviation. The NXCORR fidelity metric offers the advantage of being unaffected by the absolute position of the source centroid and the overall flux level, enabling a more equitable comparison between visibility-based and closure-only image reconstructions. We define the loss function for image reconstructions \textbf{A}' of an image \textbf{A} as the weighted sum between the NRMSE and NXCORR,
\begin{equation}
   \mathcal{L}_{\rm img}({\rm \textbf{A}, \textbf{A}'}) = \begin{bmatrix}
           \beta \\
           1-\beta
         \end{bmatrix}^\intercal \begin{bmatrix}
             {\rm NRMSE(\textbf{A}, \textbf{A}')} \\
             {1 - \rm NXCORR(\textbf{A}, \textbf{A}')}
         \end{bmatrix} \,,
\end{equation}
where the fractional weight, $\beta$, controls the contribution of each loss term. For the initial 50 training epochs, we set $\beta=1$ to prompt the machine learning model to generate centralised images. Subsequently for the following 100 epochs, the NXCORR metric is incorporated with $\beta = 0.9$, and then at $\beta = 0.5$ for another 100 epochs. We set the hyperparameters $\vec{\alpha} = (0.03, 1, 2, 3)$ in Equation \ref{eq:losses}, prioritising image reconstruction losses, particularly the product from closure invariants. This set of weights was chosen after validating the model's performance of this model on a small selection of alternative hyperparameters, and we found this set to be optimal for image reconstruction from closure invariants. A different setting (e.g. uv-coverage or field of view) may require a different set of weights to achieve ideal performance, but a full hyperparameter exploration exceeds the scope of this work. The loss from the classifier MLP is assigned a very low weight in comparison to image losses. As such, the classification loss has a negligible effect on the transformer and convolutional encoder. Throughout training, we use the \texttt{Adam} optimiser \citep{Kingma_2014} with an initial learning rate of 0.001, reducing by a factor of $0.99^n$ for $n$ spanning the number of epochs. The learning rate is reinitialised back to 0.001 whenever changes are made to the loss function. We also include a drop-out rate of 10\% in the transformer encoder to prevent over-fitting and in total, we train the model for 250 epochs. The trained model and an example notebook can be located in a GitHub repository \citep{DIReCT_Zenodo}\footnote{\href{https://github.com/samlaihei/DIReCT}{https://github.com/samlaihei/DIReCT}}.

\section{Results} \label{sec:results}

\begin{figure*}
\includegraphics[width=0.95\textwidth]{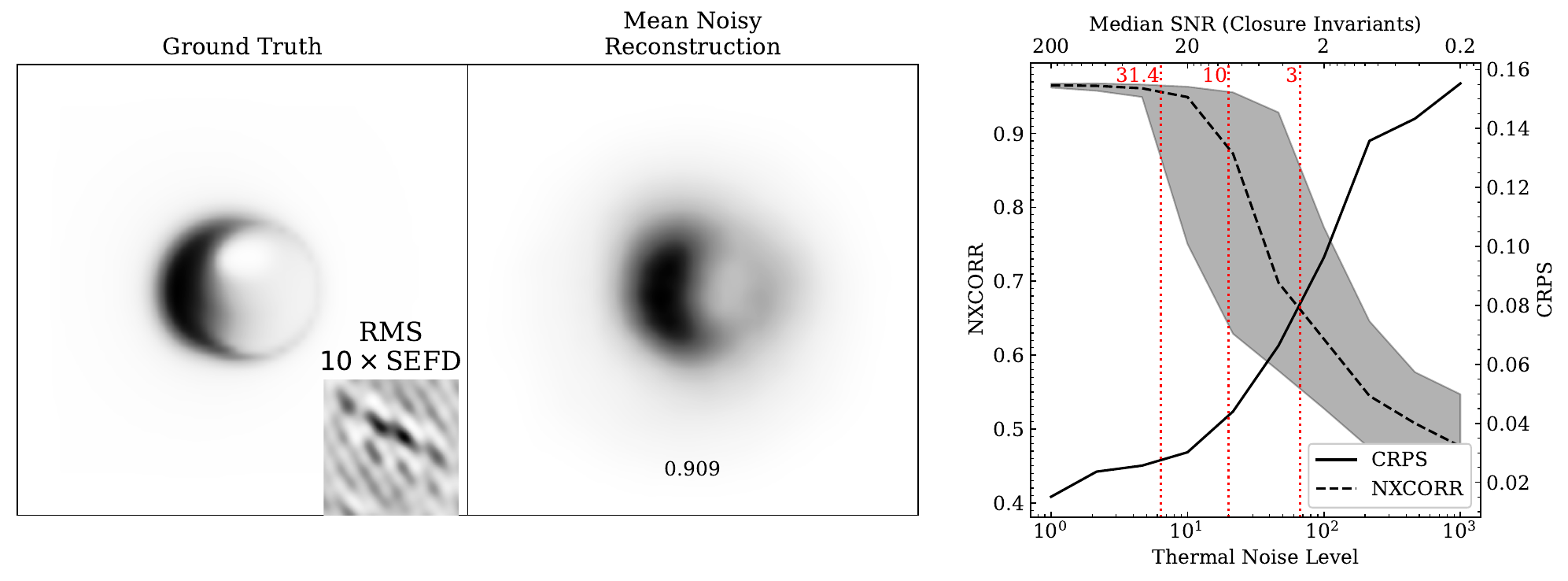}
    \caption[Thermal noise fidelity]{(Left panel) Ground truth image of a Sagittarius~A* radiatively inefficient accretion flow model \citep{Broderick_2011}. (Left inset) RMS image illustrating the correlated noise introduced when the thermal noise level is set to $10\times$ the SEFD. (Middle panel) Mean reconstruction from 1000 versions of noisy closure quantities measured from the ground truth image for $10\times$ SEFD thermal noise. The NXCORR score for the mean reconstruction compared to the noiseless ground truth is shown beneath the image. (Right panel) Relationship between NXCORR and CRPS scores to the thermal noise level and the median closure invariants SNR for the Sgr~A* example image. We show the median NXCORR as the dotted line and the sheath outlines the 16\% and 84\% percentiles of the distribution. The CRPS is plotted as the solid line. Vertical dotted lines denote three SNR thresholds: $31.4$, $10$, and $3$, which correspond to the median post-phase calibration Stokes $I$ component SNR for the primary M87 EHT dataset \citep{EHT_2019_Data}, a threshold typically used for self-calibration, and a threshold often used for low SNR flagging, respectively.}
    \label{fig:thermal-noise-fidelity}
\end{figure*}


We synthesise 230 GHz observations with the \texttt{eht-imaging} software \citep{Chael_2018_ehtim} for a static non-variable source placed at the M87* location, $\rm{RA = 12^h30^m49.42^s\, and \, Dec = +12^{\circ}23'28.04''\, (J2000)}$. The receiving bandwidth is set as 8 GHz. Our observations involve seven stations in the EHT VLBI telescope array \citep{EHT_2019_Array}: the Atacama Large Millimeter/submillimeter Array \citep[ALMA;][]{Wootten_2009_ALMA, Goddi_2019_ALMA}, the Atacama Pathfinder Experiment telescope \citep[APEX;][]{Gusten_2006_APEX}, the Large Millimeter Telescope \citep[LMT;][]{Hughes_2010_LMT}, the Pico Veleta IRAM 30 m telescope \citep[PV;][]{Greve_1995_PV}, the Submillimeter Telescope Observatory \citep[SMT;][]{Baars_1999_SMT}, the James Clerk Maxwell Telescope (JCMT), and the Submillimeter Array \citep[SMA;][]{Ho_2004_SMA}. 

An inherent assumption in our noise model is that time-dependent corruptions can be considered stable during the integration time used to measure visibilities. Therefore, we choose to synthesise observations from the array with short 5s integrations every 10 minutes over a period of 24 hours beginning on 2017-04-05 midnight UTC, $\rm{MJD} = 57848$. In total, there are 821 independent complex visibilities and 992 real-valued closure invariants, of which 912 are independent. The transformer encoder in our deep learning architecture accesses all 992 closure invariants from the full synthesis to preserve the positional encoding for each individual antenna triad, but it does not explicitly learn from the $uv$ coordinates. We illustrate the Fourier sampling of this study in Figure \ref{fig:uv-sampling} on the left panel and the corresponding dirty beam on the right panel. All images in this study are displayed with the same angular scale as the dirty beam image, which is $225\times 225\,\mu$as. The effective clean beam is a $\sim29\times20\,\mu$as ellipse.

We train the deep learning model as described in Section \ref{sec:conv_ae} and save the model with the minimum validation loss. In Figure \ref{fig:gain-noise}, we show the result of our trained deep learning model (identified as \direct) for a crescent toy model image, comparing them to \texttt{eht-imaging} reconstructions using visibilities and closure quantities for different levels of amplitude gain error. The gain errors are time-dependent station-based complex multiplicative corruptions sampled from known underlying distributions for each site (see Eq. 28 of \citealt{Chael_2018_ehtim}). The total flux of the ground truth image is normalised to 100 Jy. In this test, each reconstruction is also influenced by random, zero-mean thermal noise applied on the visibilities, which is not included during training. The complex thermal noise on the visibility is drawn from a Gaussian distribution whose width depends on the integration time, observing bandwidth, as well as each station's system equivalent flux density (SEFD) as listed in Table 1 of \citet{Chael_2018_ehtim}. On the top right panel and beneath each image, we show the NXCORR fidelity score between the reconstruction compared to the ground truth blurred by the effective beam as a function of fractional gain error. Both \direct\ and the closure-based \texttt{eht-imaging} regularised maximum-likelihood are insensitive to arbitrarily large gain errors, while the reconstructions dependent on the complex visibilities begin to fail at gain error $>25$\%. Our deep learning approach achieves a higher resolution reconstruction and comparable NXCORR with \texttt{eht-imaging} for a basic crescent test image.

To quantify the general performance of the \direct\ model under the effect of noise, particularly multi-modal posteriors, we introduce the continuous ranked probability score \citep[CRPS;][]{Hersbach_2000_CRPS}, which is a widely used metric in weather forecasting for quantifying the distance between a sample probability distribution function (PDF) and a true PDF. We define the image reconstruction CRPS as,
\begin{equation}
    {\rm CRPS(\textbf{A})} = \frac{1}{M}\sum_{i=1}^{M} \int_{-\infty}^{+\infty} [{\rm CDF}_{i}(A_r) - {\rm CDF}_{i}(A)]^2 dA\,,
\end{equation}
where ${\rm CDF}_{i}(A)$ is the pixel-wise cumulative distribution function of intensity for all reconstructions, $A_r$, of an image \textbf{A}, and ${\rm CDF}_{i}(A)$ can be represented with a Heaviside step-function centered on the true intensity of \textbf{A} at the $i$-th pixel. The squared differences in the CDFs at every pixel are summed to produce a single measure of fidelity for a sample of image reconstructions. By comparing predicted CDFs to the true intensity, the CRPS score can be used to quantify individual prediction performances, which is particularly useful where the predicted distribution exhibits multimodalities \citep{Gneiting_2007, Polsterer_2017, Polsterer_2019}. However, the CRPS is a pixel-by-pixel comparison like NRMSE. Therefore, we shift the image by the discrete value which maximises NXCORR prior to performing the calculation. As a generalisation of the mean absolute error, higher CRPS values indicate a greater departure in the intensity distribution of the reconstruction compared to the truth image. 

Although multiplicative corruptions are canceled out from the closure quantities to zeroth order, the error terms are scaled with the thermal noise level to first order and higher. We aim to quantify the performance of our network under the effect of different levels of thermal noise. To perform this test, we set the fractional gain error budget to 10\%. We have chosen a 10\% gain error because self-calibrated residual gain corrections can vary by over an order of magnitude between $3-30\%$, but the majority of gain corrections fall below 3\% \citep[e.g.][]{Chael_2018_ehtim}. We then generate 1000 noisy sets of visibilities for each level of thermal noise error from $1-1,000$ times the estimated thermal noise contribution based on each individual station's SEFD. For every set of visibilities, we compute the resulting closure invariants and create the reconstruction with the trained model. Each individual noise level produces an NXCORR distribution of all 1000 noisy reconstructions and one CRPS summary statistic. In Figure \ref{fig:thermal-noise-fidelity}, we illustrate the relationship between the reconstruction fidelity metrics with the thermal noise level for an untrained radiatively inefficient accretion flow model of Sagittarius~A* \citep{Broderick_2011}, shown in the left panel of the figure. The total flux of the image is normalised to 100 Jy. The left inset shows the RMS image, illustrating the correlated noise introduced when the thermal noise level is $10\times$ SEFD. The middle panel displays the mean \direct\ reconstruction from closure invariants where the $10\times$ SEFD thermal noise has been applied to visibilities, and the right panel plots the CRPS and NXCORR metrics as functions of the thermal noise level. The median NXCORR is displayed with the dotted line and the sheath captures the central 68\% of the distribution. The image fidelity remains consistently at $\gtrsim 0.95$ until the thermal noise is $\gtrsim 10$ times the estimated noise contribution based on each station's SEFD. Subsequently, the accuracy of the reconstruction declines rapidly. 

As the SEFD values are highly specific to the array properties, we introduce a signal-to-noise ratio (SNR) of the closure invariants. For each level of thermal noise, we compute the SNR distribution of the independent set of closure invariants with the standard deviation from 1000 noisy visibilities. With the exception of very low SNR, the median of the distribution is approximately related to the thermal noise level by a simple power-law and we show the median SNR of closure invariants associated with the Sgr~A* image in the secondary $x$-axis of Figure~\ref{fig:thermal-noise-fidelity}. We also indicate three SNR thresholds: $31.4$, $10$, and $3$, corresponding to the median post-phase calibration Stokes $I$ component SNR for the primary M87 EHT dataset \citep{EHT_2019_Data}, a threshold typically used for self-calibration, and a threshold often used for low SNR flagging, respectively. We further emphasise that closure invariants, as combinations of Fourier components, are more affected by thermal noise than complex visibilities. While the numerical values shown in Figure \ref{fig:thermal-noise-fidelity} (right panel) are specific to the reconstruction of the Sgr~A* ground truth image, reconstruction fidelity of diverse morphological classes show a consistent general relationship with noise. 

Though closure invariants do not constrain the total flux of the image, the impact of thermal noise on closure invariants is influenced by the integrated flux. When the source is normalised to 100 Jy, the closure invariants are robust to thermal noise at $1\times$ SEFD, but for fainter sources on the order of 1 Jy, as is the case for M87* and Sgr~A*, thermal noise on the closure invariants can produce unreliable predictions. In forthcoming studies, incorporating noisy variations of closure invariants will assist the network in generating predictions that are even more resilient to thermal noise, enabling more robust predictions for fainter targets.

\subsection{Non-horizon source structures}
\begin{figure*}
\begin{tabular}{cc}
  \includegraphics[width=84mm]{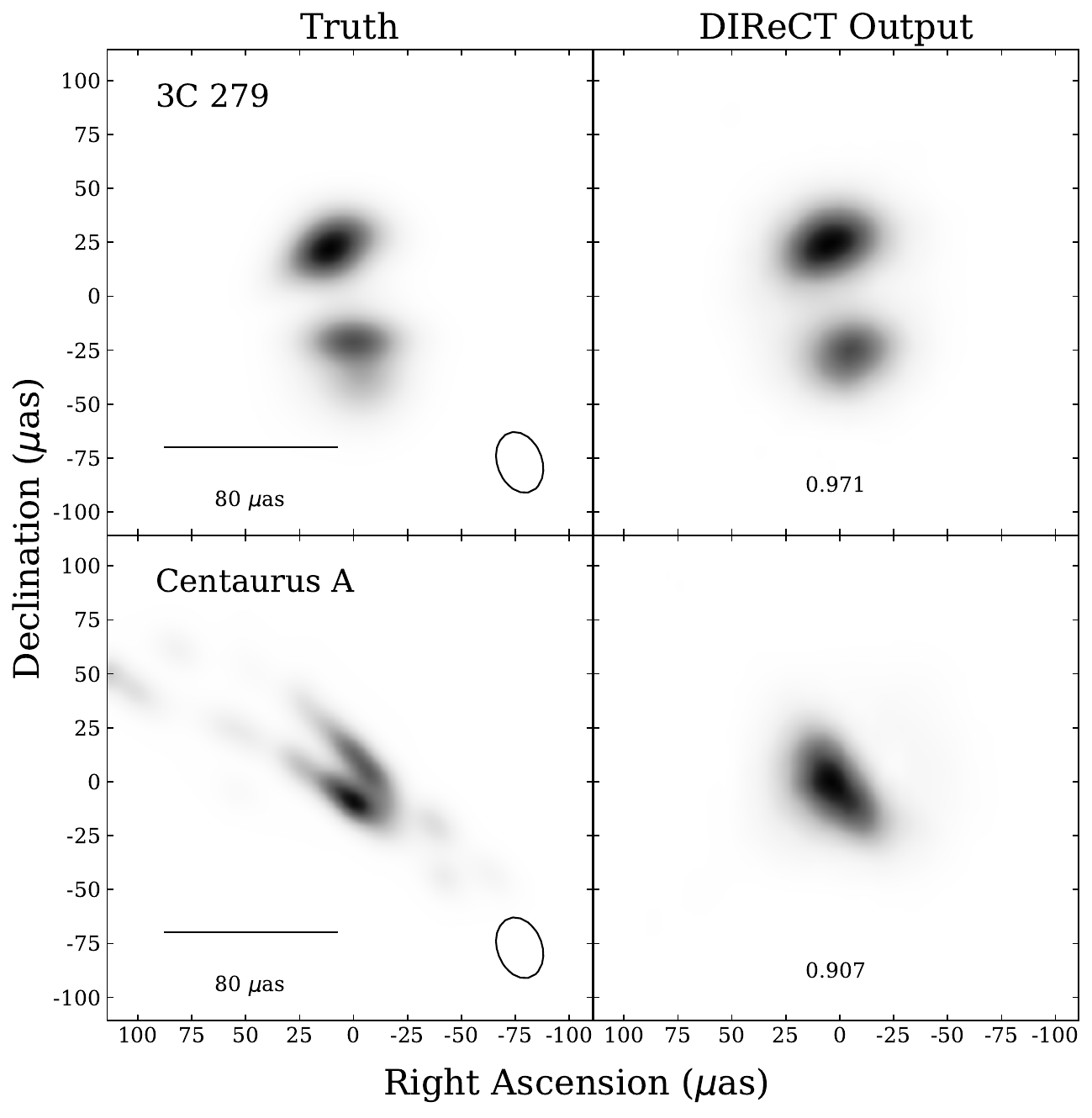} &   \includegraphics[width=86mm]{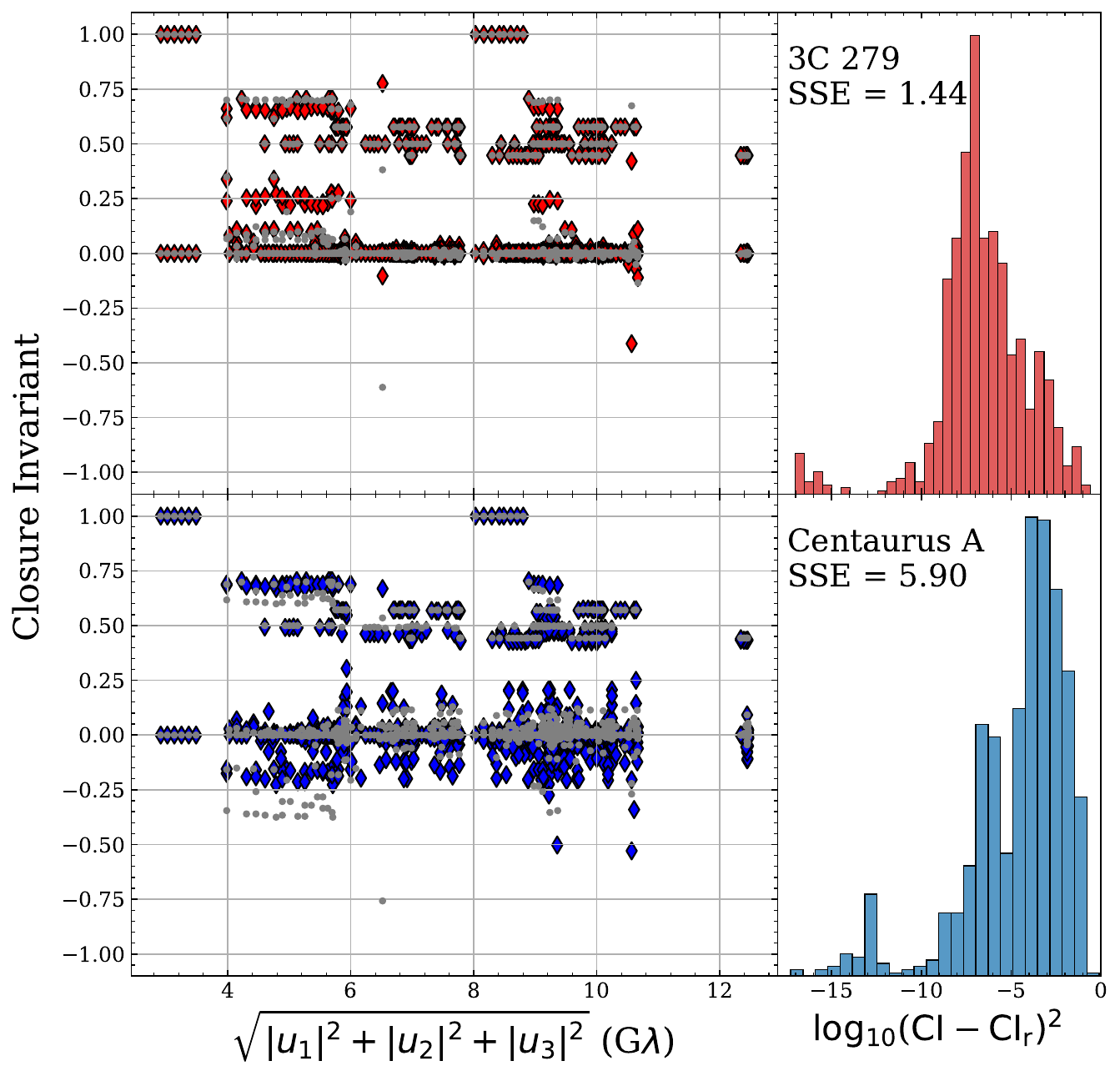} \\
\end{tabular}
\caption{(Left) Models of 3C279 (top) and Centaurus A (bottom) seen within the field of view of the \direct\ model ($225\times225\,\mu$as). The size scale and he effective beam size of our $uv$-coverage are shown in the ground truth panel. The image reconstructed from synthetic observations is presented in the middle panels labeled as ``\direct\ Output''. NXCORR fidelity scores relative to the unblurred ground truth, are labeled below. (Right) Illustrations of the fits between the ground truth closure invariants, shown as filled diamonds, and closure invariants from the reconstruction, overplotted as points. On the left panels corresponding to 3C 279 (top) and Centaurus A (bottom), the magnitudes of the real-valued closure invariants are plotted relative to the sum of squared triad baseline lengths. On the right panels, we present a histogram of the squared errors in logarithmic scale and the sum of squared errors between the two sets of closure invariants is shown in the top left corner.} \label{fig:astroSources}
\end{figure*}

It is not uncommon for the data obtained through VLBI observations to be modelled geometrically with the linear combination of simpler primitive shapes \citep[e.g.][]{Cornwell_2008_CLEAN, EHT_2019_ShadowMass, Broderick_2020_Themis, Roelofs_2023_mring}. A prominent example is the blazar 3C 279, which was imaged with the Event Horizon Telescope at $\sim20\,\mu$as resolution \citep{Kim_2020_3C279}. The resulting image can be robustly modelled with six Gaussian components across four separate epochs. Here, we demonstrate that the core-jet structure of 3C 279 can be recovered by the \direct\ approach. We obtain the Gaussian decomposed model from Table D1 of \citet{Kim_2020_3C279}, rescale it to the field of view of our model ($225\times225\,\mu$as), and simulate observations with the EHT array. The core-jet separation in the rescaled image is $\sim50\,\mu$as, in contrast to $\sim75\,\mu$as in the original field of view. We show the multi-Gaussian model and the \direct\ reconstruction in the top two panels of Figure \ref{fig:astroSources} (left). The \direct\ reconstruction reconstructs the brightest two components of 3C 279, but it is prejudiced by augmented doubles in the training data, as evidenced by the fact that both components in the reconstruction are sheared in the same direction unlike the components in the ground truth which are oriented differently. Hence applying augmentations to individual components in training data prior to their combination in the image space would further enhance the model's performance on general complex morphologies resembling multiple Gaussian components. Nevertheless, the surface brightness ratio, core-jet separation, orientation, and modest asymmetry of the brightest component are faithfully reproduced.

In contrast to 3C 279, Centaurus A is an independent source imaged by the Event Horizon Telescope \citep{Janssen_2021_CenA}, showing edge-brightened filamentary structures which are not as easily decomposed into a sparse set of Gaussian components. Moreover, the source morphology is structurally distinct from the shapes in \direct's training set, representing a useful test of the \direct\ model's generalisation capabilities. We obtain the final model image corresponding to Fig. 2 of \citet{Janssen_2021_CenA} and rescale it to our model's field of view. We then apply the \direct\ pipeline, presenting the ground truth and reconstructed results in the bottom two panels of Figure \ref{fig:astroSources} (left). As the Centaurus A model is placed in the sky of location of M87, which is inaccessible by the South Pole Telescope \citep[SPT;][]{Carlstrom_2011_SPT, Kim_2018_SPT}, the loss of the SPT baselines results in a degradation of the fine detail of Centaurus A's jet launching region and collimation profile. These angular resolution limitations cause the low surface brightness counter-jet and the filamentary ridge structures in the downstream jet region to be effectively unresolved. The resulting \direct\ reconstruction resembles a necktie knot that reproduces the asymmetry and orientation of the edge-brightened jet nearest to the jet apex.

On the right-hand side of Figure \ref{fig:astroSources}, we show the complete set of closure invariants measured from the ground truth as filled diamonds and from the reconstruction as points for the reconstructions of 3C 279 and Centaurus A. In the left panels, the magnitudes of the real-valued closure invariants are plotted relative to the sum of the squared triad baseline lengths, where $u_s$ is the baseline length of station, $s$, which constitutes one leg of the triad. Although the model was not explicitly optimized to minimize the deviation between observed and reconstructed closure invariants $(\rm{CI} - \rm{CI}_r)^2$, it demonstrates satisfactory performance in fitting the data terms. On the right panels, we plot the distribution of squared errors in logarithmic scale and quantify the overall fit performance with the sum of squared errors, $\rm{SSE} \equiv \sum(\rm{CI} - \rm{CI}_r)^2$, shown on the top left corner of the histogram panel. Although not fit to the interferometric data explicitly, \direct's output is often an accurate reconstruction of the source morphology, making it particularly well-suited for initialising forward-modelling minimisation algorithms as an image prior for further refinement. This extra refinement step would enforce consistency between the final image and its interferometric observables. In future works, we will incorporate closure invariant minimisation as a post-processing step in the \direct\ reconstruction pipeline.

\subsection{Performance on untrained general morphologies}

\begin{figure*}
\begin{tabular}{cc}
  \includegraphics[width=60mm]{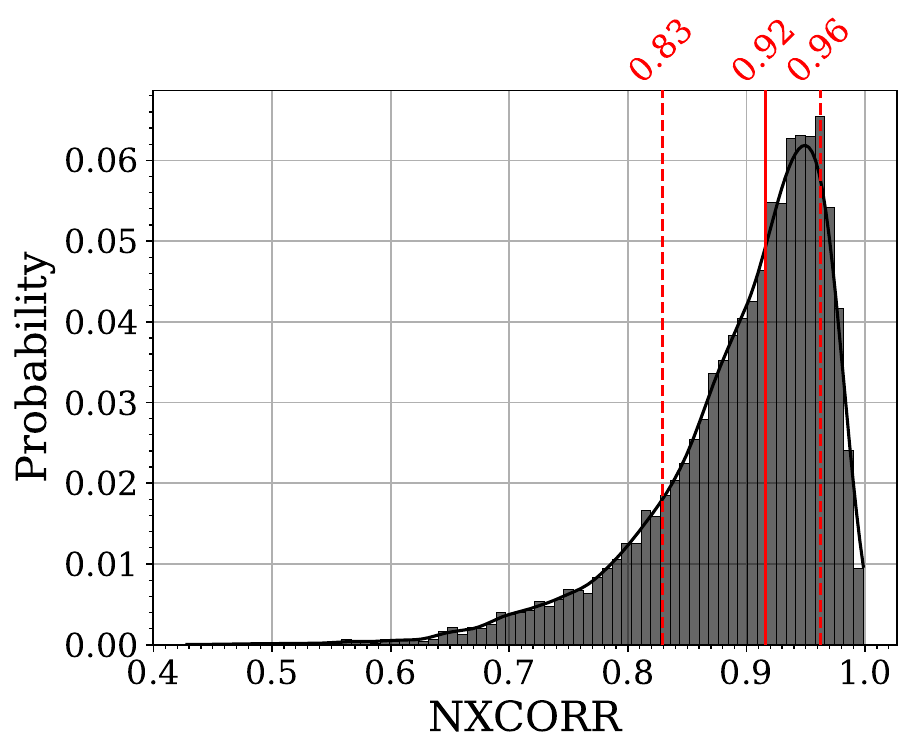} &   \includegraphics[width=115mm]{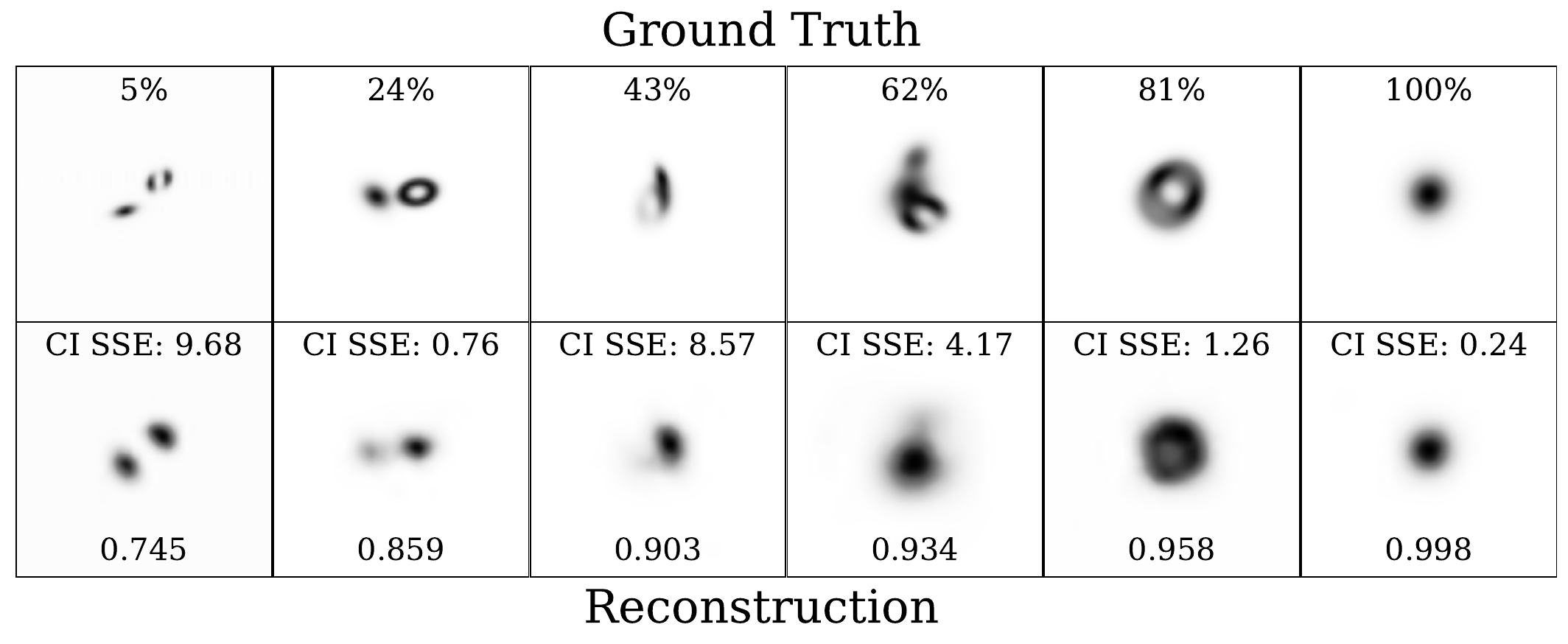} \\
\end{tabular}
\caption{(Left) NXCORR probability distribution for image reconstructions of 10,000 random augmented image combinations representing untrained source morphologies. The 16\% and 84\% percentiles are marked by the dashed line and the median by the solid line with their corresponding fidelity scores are displayed above the plot. (Right) Panels present the ground truth image (top) and reconstruction (bottom). We show ground truth and reconstructed images corresponding to six equally-spaced NXCORR percentiles bins spanning the 5\% and the 100\% percentiles. We also present the sum of squared errors between the set of ground truth and reconstructed closure invariants.} \label{fig:mixed-combinations-fidelity}
\end{figure*}

In this section, we estimate the performance of the \direct\ model on untrained morphologies by constructing 10,000 random augmented image pairs from simple models of gausses, ellipses, rings, $m$-rings, and doubles in the validation dataset. Each image prior to combination is subject to rotation, up to $30^{\circ}$ shear, up to 20\% resizing, and translation by up to 20\% the image width. We highlight that the 30\% shear and image translation are entirely unseen by the network, given that the training data are augmented with up to $5^{\circ}$ shear and training images are always centered. These untrained augmentations test the network's ability to generalise. For each combination image, we measure the NXCORR between the deep learning reconstruction and the ground truth. We show the distribution of NXCORR values for all 10,000 random combination images in Figure \ref{fig:mixed-combinations-fidelity}. The median NXCORR fidelity is marked by the solid line and the dashed lines denote the 16th or 84th percentile reconstructions. The values corresponding to each percentile are shown above each vertical line and right panels present example combination images representing linearly-spaced six percentile bins spanning 5th to 100th percentiles. Each ground truth combination image plotted above is accompanied by the corresponding reconstruction and NXCORR fidelity metric annotated below. For this test, we aim to quantify the reconstruction fidelity for general untrained morphologies. As such, we do not apply thermal noise to the visibilities. 

Figure \ref{fig:mixed-combinations-fidelity} shows a median reconstruction NXCORR fidelity of $\simeq 0.92$ is achieved, but the distribution is skewed left with outliers reaching $\rm{NXCORR}\simeq 0.4$ and the mode of the distribution is $\gtrsim 0.9$, near $\sim0.95$. Qualitatively, the sizes and relative orientation of components in all six example combination images shown on the right panels are faithfully reproduced in the reconstruction, even for the 5th percentile result which achieved $\rm{NXCORR}\simeq 0.75$. The SSE in the measured closure invariants between the ground truth and reconstruction is shown alongside each reconstruction, but we caution against comparing SSEs between reconstructions with different ground truth images. Due to the high feature complexity of the combination images, the multi-layer perceptron classifier is most likely to predict the CIFAR-10 class from the latent features instead of a mixed combination of the constituents. This experiment demonstrates that the trained model achieves good performance even for untrained source morphologies consisting of mixed pair combinations of simpler classes.

\section{Discussion} \label{sec:discussion}
The principal motivation behind this work was to demonstrate a proof-of-concept that the co-polar closure invariants formalism outlined in \citep{Thyagarajan_2022_CI} can be used to directly reconstruct images. We achieve this with the \direct\ approach, by training a deep learning network which can extract features from interrelationships from the set of closure invariants using the attention mechanism of the transformer, and we show that the network succeeds in image reconstruction under a very challenging regime, with the extreme sparsity of the Event Horizon Telescope $uv$-coverage. Furthermore, the result is robust against multiplicative station-based corruptions and additive thermal noise, and the network can be generalised for reasonable performance on untrained morphologies. 

A critical advantage in the \direct\ approach to image reconstruction is that once the machine learning model is trained, there are no additional hyperparameters to hand-tune based on a calibration dataset and no explicit prior or regularisation terms used to influence each individual reconstruction. Instead, the training dataset, composed of natural images of the CIFAR-10 sample and simple mathematical shapes, forms the implicit prior for the model. The training dataset itself presents several notable biases. For instance, the radial Gaussian tapering applied on the CIFAR-10 dataset biases the reconstruction toward centralised emission sources, particularly sources where the peak emission is in the central pixel. The Gaussian, ellipse, disk, and double classes further reinforce this bias. By convolving training images with a Gaussian kernel, we also train the model to produce smoothly varying surface brightness distributions.

The majority of our training dataset is composed of CIFAR-10 natural images. \citet{Feng_2024} notes that the CIFAR-10 dataset as a prior is preferentially biased for purely horizontally or vertically oriented edges, resulting in boxy artifacts appearing in place of circular image features. We did not encounter boxy artifacts in our model likely as a result of our rotational and shearing augmentations during training, which randomises the edge orientation.

\begin{figure*}
	\includegraphics[width=1.0\textwidth]{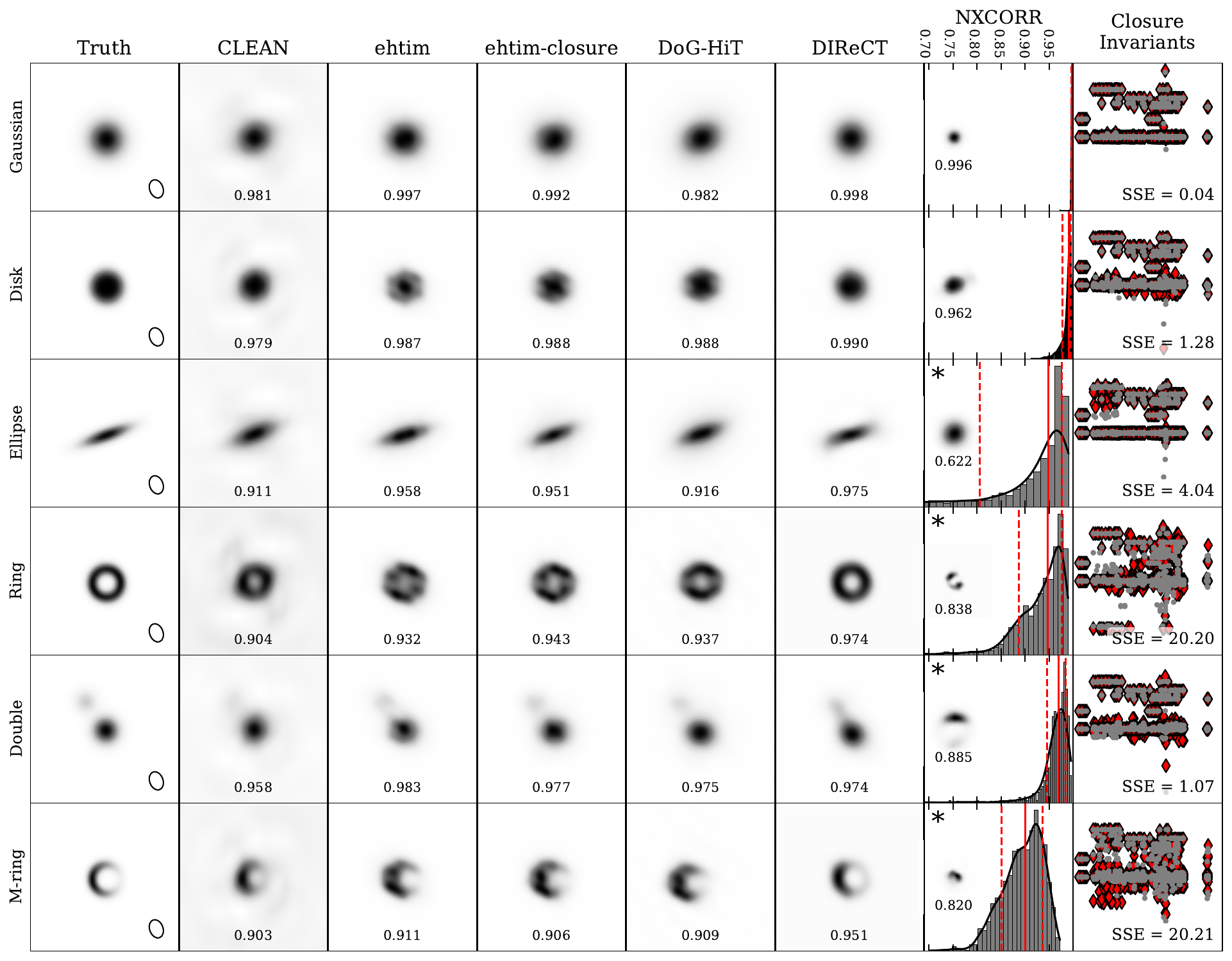}
    \caption[Matrix image reconstruction comparison]{Image reconstructions of sample images representing distinct morphological classes of the training dataset with state-of-the-art methods (standard \texttt{CLEAN}, \texttt{eht-imaging}, \texttt{DoG-HiT}) compared to our trained machine learning model, \direct. We show the effective beam dimensions in the ground truth panels. NXCORR fidelity scores are underneath each reconstruction. In the penultimate column, we show the NXCORR distribution for reconstructions of the truth image under a variety of augmentations which include rotation, rescaling, and up to $30^{\circ}$ shearing (if asterisk is included). Vertical lines indicate 16th, 50th, and 84th percentiles. Inset images in the fidelity distribution panel show the 5th percentile reconstruction as examples of potentially suboptimal reconstructions, but their corresponding NXCORR fidelity scores indicate that even the 5th percentile reconstructions are reasonably good. In the final ``Closure Invariants'' column, we present the fits between the ground truth (diamonds) and reconstructed (points) closure invariants alongside their sum of squared errors. The dimensions of the closure invariants panels are identical to Figure \ref{fig:astroSources}.}
    \label{fig:recon-comparison}
\end{figure*}

As closure quantities capture less information on the source structure and their uncertainties are more strongly influenced by thermal noises than the full complex visibilities, one potential source of concern for closure-based imaging is the dynamic range of the reconstruction \citep[e.g.][]{Akiyama_2017_imaging, Chael_2018_ehtim}. Closure-based maximum likelihood methods found spurious low-luminosity artifacts in the image reconstruction, resulting from local minima in the objective function \citep[e.g.][]{Chael_2018_ehtim}. However, the image reconstruction method presented in this study represents a different approach. Notably, our results are resistant to noise in both the image and Fourier space, and the implicit imaging priors from the training dataset discourages the appearance of low-luminosity artifacts. Nevertheless, the dynamic range achievable by any reconstruction algorithm on a given set of interferometric measurements is dependent on the noise characteristics of the array and our reconstruction is likewise subject to these limitations. We leave a full discussion of the dynamic range in the array and our reconstruction for a future study where our model is applied to actual VLBI data.

\begin{figure*}
	\includegraphics[width=1.0\textwidth]{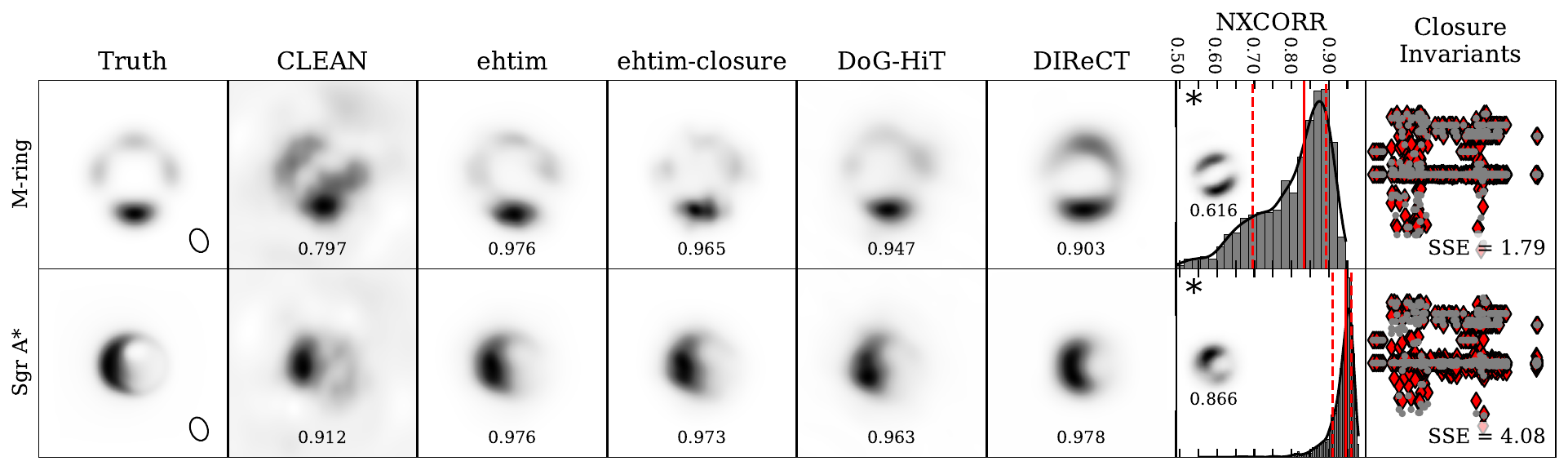}
    \caption[Matrix image reconstruction comparison]{Same construction as in Figure \ref{fig:recon-comparison}, but the sample images from the training dataset are replaced with a fourth order $m$-ring and a Sagittarius~A* model \citep{Broderick_2011}, which are untrained morphological classes.}
    \label{fig:recon-comparison_untrained}
\end{figure*}

\subsection{Comparing Reconstructions}
\begin{figure}
	\includegraphics[width=0.9\columnwidth]{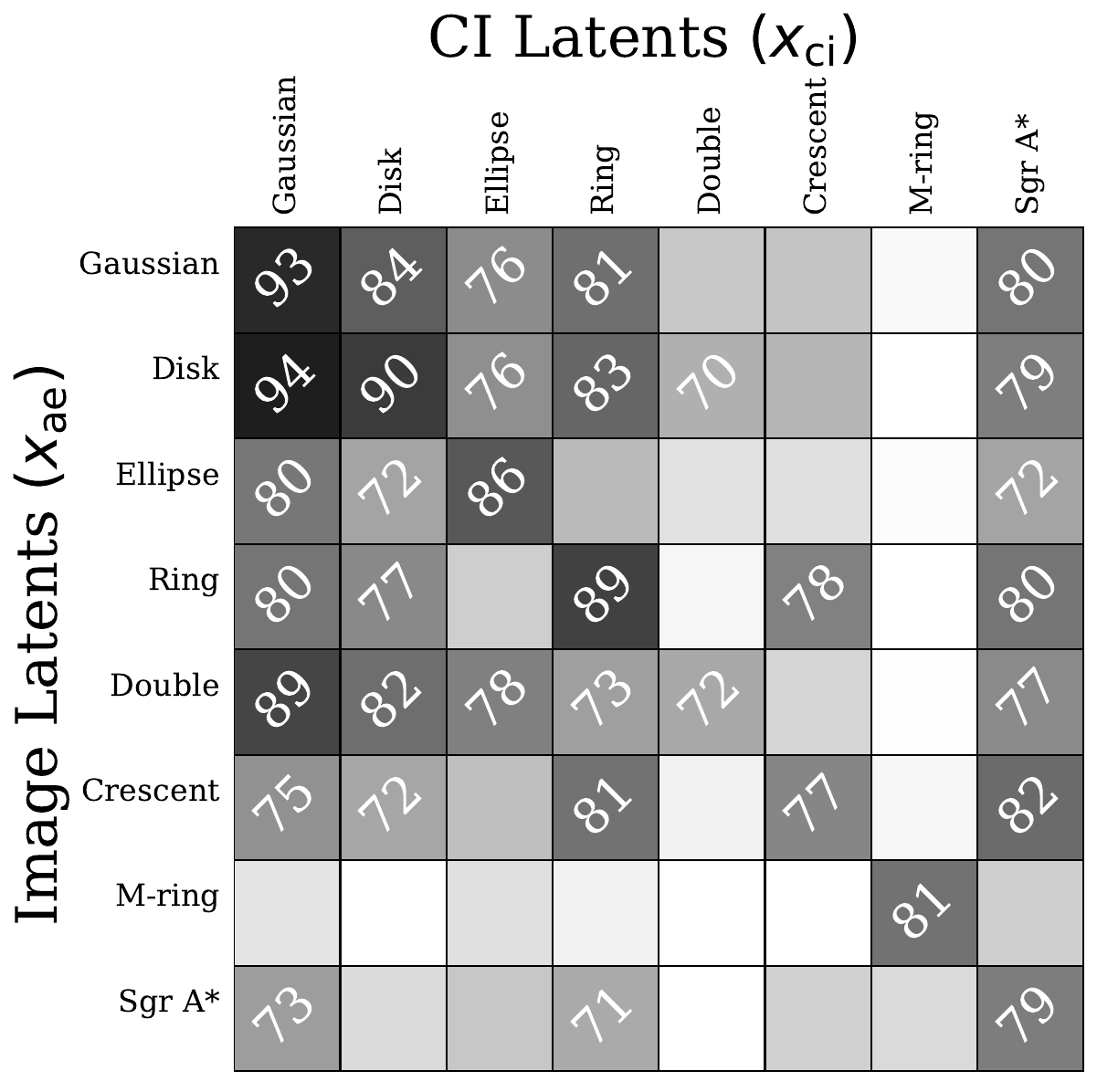}
    \caption[]{Pair-wise cosine comparison matrix of latent encoded features generated from closure invariants ($x_{\rm{ci}}$) and images ($x_{\rm{ae}}$) for both trained and untrained morphological classes in Figures \ref{fig:recon-comparison} and \ref{fig:recon-comparison_untrained}, respectively. Similarity scores are distinguished by shade and scores above 70\% are labeled.}
    \label{fig:feature-similarity}
\end{figure}

In Figures \ref{fig:recon-comparison} and \ref{fig:recon-comparison_untrained}, we present reconstructions from our trained machine learning model, \direct, of ground truth morphologies representative of trained and untrained classes, respectively. We compare the result with standard \texttt{CLEAN} and regularised maximum-likelihood methods (\texttt{eht-imaging} and \texttt{DoG-HiT}) described in Section \ref{sec:imrec}. All reconstructions are created from noiseless synthetic observations with an identical VLBI array and regularised maximum-likelihood methods are initialised with an identical Gaussian prior. Each row represents one ground truth model and each reconstruction method is distinguished by column, with \texttt{eht-imaging} separated into reconstructions based on visibility (ehtim) or closure (ehtim-closure) data terms. 
The NXCORR value for each reconstruction is annotated on each panel. The penultimate column quantifies the NXCORR fidelity distribution of \direct\ for the ground truth image class under a variety of image augmentations which include rotations, rescaling, and up to $30^{\circ}$ shear where the asterisk is shown. Vertical lines denote the 16th, 50th, and 84th percentiles as in Figure \ref{fig:mixed-combinations-fidelity}. We also present the 5th percentile reconstruction from the distribution in the inset image on the fidelity distribution panel as an example of reconstruction with potentially suboptimal fidelity. However, the corresponding NXCORR values indicate that even the 5th percentile reconstructions have reasonable fidelity. Like Figure \ref{fig:astroSources}, we plot the fits between the ground truth, represented as filled diamonds, and reconstructed closure invariants, represented as points, alongside the SSE between the two sets of measurements in the final column. The closure invariants are plotted relative to the triad baseline lengths on the same scale as Figure \ref{fig:astroSources}.

For Figure \ref{fig:recon-comparison}, we present six of the distinct morphological classes (gauss, disk, ellipse, ring, double, and $m$-ring) from the validation dataset. All of the image reconstruction algorithms achieve $>0.9$ NXCORR fidelity in their reconstruction output with the two variants of \texttt{eht-imaging} performing marginally better in most cases. Due to the convolution with the effective beam, most reconstructions appear as blurred versions of the ground truth, as is most evident in the ring thickness for the ring and $m$-ring classes. For the \direct\ output, we generally achieve better clarity in our image than other algorithms, resulting in a sharper and more defined central depression for the ring class. The closure invariants column indicates that large residuals on the closure data terms does not necessarily indicate a lower quality fit in the image space. When augmentations are considered, the rotationally invariant gauss and disk morphologies are reliably reproduced with near absolute NXCORR fidelity across the range of rescaling augmentations, but the NXCORR declines to a median value of $\simeq 0.95$ when the axisymmetry is disrupted as in ellipse sources. 

\direct's reconstruction fidelity of ring structures is similar to ellipses, but declines to $\simeq 0.9$ for a first order $m$-ring, seen as a crescent structure. With single Gaussian and disk excepted, the disjointed double disk structure is more faithfully reconstructed under augmentation than all other trained classes. Moreover, the 5th percentile reconstruction, which serves as an example of a potentially suboptimal reconstruction, is typically a reasonable representation of an augmented version of the ground truth with some notable exceptions, which are the result of a classification error. The inset panel images for the double and the ring class both appear as second order $m$-rings and the ellipse is reconstructed as a Gaussian. In many cases, the 5th percentile reconstruction, which we present as a proxy for a potential poor reconstruction, is often a reliable reconstruction with NXCORR $\gtrsim 0.8$. 

Of all of the reconstruction methods demonstrated here, the \texttt{CLEAN} algorithm is most likely to produce spurious features associated with the sidelobes of the dirty beam, but this issue can be identified and remedied by tuning the iterative deconvolution procedure. We also note that the surface brightness distributions in the ring reconstruction for both the \texttt{eht-imaging} and \texttt{DoG-HiT} algorithms are not uniform. Likewise, artifacts appear at the edges of the disk reconstruction. The regularised maximum-likelihood methods can produce subtle super-resolution artifacts, which can be further mitigated by convolving with the complete effective beam size (as opposed to half of the effective beam dimensions) to prevent over-interpretation. 

In Figure \ref{fig:recon-comparison_untrained}, we show a comparison of the image reconstruction methods for a fourth order $m$-ring and the Sgr~A* model \citet{Broderick_2011} shown in Figure \ref{fig:thermal-noise-fidelity} as examples of untrained classes. The finer details of each image reconstruction for the fourth order $m$-ring vary significantly between independent algorithms. Notably, \texttt{CLEAN} captures some of the structural complexity, but not the dynamic range in the surface brightness distribution. Similarly, the \direct\ model's reconstruction captures some of the secondary structure, but variations in the surface brightness and the specific locations of secondary hot spots are not accurately reproduced. The augmentation fidelity distribution is skewed left with a median of $\simeq 0.84$. The 5th percentile reconstruction (NXCORR $\approx0.62$) in the $m$-ring NXCORR fidelity distribution shares similar characteristics to the \texttt{CLEAN} reconstruction in that the size of the ring reconstruction is accurately reproduced, but the surface brightness distribution does not correspond to the hot spots in the ground truth. The two variants of \texttt{eht-imaging} and \texttt{DoG-HiT} successfully reconstruct the four hot spots, but the brightness distribution is most accurate in the \texttt{eht-imaging} reconstruction which has access to the full error-free complex visibilities. 

As for the Sagittarius~A* model, the \texttt{CLEAN} reconstruction produces spurious features, while the \texttt{eht-imaging} and \texttt{DoG-HiT} reconstructions appear appropriately as edge-brightened disks with a depression in the upper half. The output from the trained network is similar to that of a crescent, where the depression feature is centralised. As a whole, the reconstruction fidelity distribution of the \direct\ model under augmentation is similar to that of the trained double class. We note that despite the lower overall NXCORR fidelity score, even the 5th percentile reconstruction with $\rm{NXCORR } \approx 0.87$ reproduces an offset depression as expected from the ground truth model. 

In our evaluations, \direct\ outperforms \texttt{CLEAN} and its performance when tested against regularised maximum-likelihood methods is comparable even when handling complex untrained morphologies, which highlights the network's capacity to generalise for untrained data. For source morphologies that are similar to sources in the training dataset, \direct\ produces sharper images with fewer artifacts than the regularised maximum-likelihood reconstructions. Another key advantage lies in the time complexity of generating image reconstructions. Once trained, the machine learning network can produce the reconstruction efficiently, enabling rapid and precise image predictions that are conducive to downstream tasks, such as parameter inference and posterior estimation. Considering the model's performance and reliability in distinguishing and reconstructing sources resembling those in its training dataset, it's clear that expanding the data volume would further enhance the network's adaptability to diverse scenarios. 

An important stage of image analysis, especially as it pertains to VLBI images of black holes, is the extraction of ring features, characterising the geometry and intensity profile of the reconstructed image with a sparse set of parameters \citep[e.g.][]{EHT_2019_Imaging, Carilli_2022}. These parameters generally include the ring intensity, diameter, width, orientation, and asymmetry. The ring diameter is of particular interest as a test of general relativity and an estimate of the supermassive black hole mass. In this work, we present a proof-of-concept for direct image reconstruction from closure invariants using a deep learning approach and we leave the feature parameterisation for a future work where we test the trained model on synthetic and real data published by the Event Horizon Telescope Collaboration. Although the image reconstruction with closure invariants alone shows promise, utilising corrupted visibilities as an additional input for the transformer can assist in model convergence and resolve the inherent degeneracy in closure invariants concerning the absolute position and flux of the source. Another extension would involve incorporating the array configuration into the model, facilitating generalisation to a diversity of observations. However, this enhancement would necessitate a significant increase in training efforts from data volume to training duration.

\subsection{Similarity of Latent Features}
Figure \ref{fig:feature-similarity} provides further insight on the performance of the network by illustrating the pair-wise cosine similarity between the latent features generated by the closure invariants through the transformer ($x_{\rm{ci}}$) and by the image through the convolutional encoder ($x_{\rm{ci}}$). The cosine similarity is the dot product between the two sets of normalised vectors and expresses similarity between two vectors by the angle between them. In Figure \ref{fig:feature-similarity}, cosine similarity scores are distinguished by shade, but only scores $>70\%$ are labeled. We show all of the example images shown in Figures \ref{fig:recon-comparison} and \ref{fig:recon-comparison_untrained} representing trained and untrained morphological classes, respectively. We label the first-order $m$-ring in Figure~\ref{fig:recon-comparison} as a crescent to differentiate it from the fourth order untrained $m$-ring. 

Throughout training, the adopted mean squared error loss function between the two sets of latents promotes maximising the cosine similarity of the diagonal elements in the trained subset. While diagonal elements typically show the highest similarities, significant off-diagonal scores indicate strong cross-class feature correspondences, notably between rings/crescents and Gaussians/disks, suggesting shared morphological characteristics. The columns for the untrained classes ($m$-ring, Sgr~A*) show two patently different outcomes when reconstructing sources unrepresented in the training dataset. It is evident that the latent representation of the fourth order $m$-ring shares minimal resemblance to any of the other trained analytical classes. The low degree of correspondence with other classes suggests that the closure invariants provide sufficient information to differentiate $m$-rings from alternative structural forms with high specificity. The observed latent space differentiation gives confidence that the model, if trained on higher-order $m$-rings, would likely demonstrate high reconstruction fidelity. 

The case of Sgr~A* is inverted in that, except for the $m$-ring, the latent features share high similarity with all other classes. However, the only class with $>80\%$ similarity in latent space is the crescent which is reflected in the image reconstruction despite the similarity being only marginally higher than other classes. In a future study, we will explore a contrastive learning approach, which simultaneously maximises similarity in the diagonal elements and penalises off-diagonal similarity. Although the latent features are degenerate, we can utilise multiple different loss functions and apply them at appropriate stages during training to ensure that the model converges to a sensible solution. 

\subsection{Applications and Limitations}
The \direct\ model described in this study is trained noiselessly on a fixed VLBI array of identical stations, fixed pointing, time, field of view, and pixel scale. While these properties underscore the limitations of this study's model, many of these shortcomings can be addressed in future upgrades. Despite the challenges, this model is already well-suited for monitoring survey applications, such as the Monitoring Of Jets in Active galactic nuclei with VLBA Experiments \citep[\texttt{MOJAVE};][]{Lister_2018} project, where a fixed patch of sky is monitored by the same array over multiple epochs. Rather than predicting the image from each epoch independently, the \direct\ model can be modified such that the processing of measurements from a new epoch can be conditioned on the image from the previous epoch. The \texttt{MOJAVE} project offers a wealth of interferometric data on astrophysically relevant sources on which the \direct\ model could be trained and the \direct\ approach would offer the \texttt{MOJAVE} project a rapid and independent imaging technique for new observations.

While the long-term \texttt{MOJAVE} program is suited for observing snapshots of relatively slowly varying sources, some sources, such as Sgr A* \citep{Wielgus_2022}, vary on much shorter timescales. In principle, the \direct\ model can produce predictions from a subset of its total input data, allowing one to partition the trained 24-hour observation period into smaller chunks for visualisation. However, the model as described is not designed for this purpose and the fidelity of the result is strongly contingent on the distribution of the transformer's attention weights. Instead, time-dynamic image reconstructions would benefit most from techniques which can correlate visibility or closure terms across time, allowing all of the data to be used in a single forward-pass to reconstruct robust dynamical images. For this task, an autoencoder with a sparse inner layer can, in principle, characterise the shared information between different frames with a low-dimensional space. This is yet another promising research direction and application for future versions of the \direct\ approach.

If the interferometric measurements are obtained with a different array, if reconstructions on a field of view larger than $225\times225\,\mu$as are desired, or if higher resolution is desired, the entire \direct\ model, not merely the transformer attachment, would have to be retrained, because the latent features are optimised to facilitate their prediction from the closure invariants input. The total wall-clock time required to retrain the model using the strategy described in Section \ref{sec:training_network} is 2.5 days on a commercially available NVIDIA RTX A6000 GPU, which is significantly longer than a regularised maximum-likelihood method would require to reconstruct an image with the same data. Therefore, this design is best utilised when the trained model can be used repeatedly, where each prediction task can be completed in a fraction of a second. With this model, we empirically estimate that scaling the model to accept a larger volume of data, \textit{i.e.} by a factor of $n$, would require a factor of $\mathcal{O}(n^{0.5})$ increase in time complexity in the closure invariant computation, which makes the model suited for extension into multiwavelength VLBI and polarimetric imaging. Increasing the pixel dimensions by a factor of $n$ increases the wall-clock time in a complex and non-linear fashion due to changes that need to be propagated throughout the machine learning architecture; however we can estimate that the loss calculation applied once at every epoch, costs $\mathcal{O}(n^2)$ more time as expected for square images. But the loss calculation is a negligible fraction of the total wall-clock time required to train the model.

Future applications of \direct\ would also benefit substantially from realistic modelling of interferometric noise. As noted in \citet{Lockhart_2022}, station gain error terms can not be fully detached from the closure likelihood function. Therefore, model-fitting approaches must assume $\mathcal{V}'_{ab} = g_a \mathcal{V}_{ab} g_b^* + \epsilon_{ab} \approx V_{ab}$ in the error on closure terms to preserve their independence from gain error in the likelihood function. However, the actual likelihood may be misrepresented when the gain terms are large. The \direct\ approach offers an alternative where noise augmentations can be applied to the training data for the network to model and neither visibility nor closure terms are used in the loss function. The inclusion of realistic data corruptions also distances our approach from its original assumption of a homogeneous array. By applying corruptions based on the characteristics of each individual station, the network would be able to learn the optimal weighting distribution for image reconstruction. Implementing specific time-frame windows and individual station data masks during the training phase would mitigate the risk of over-reliance on individual, highly sensitive stations (\textit{e.g.,} ALMA in the EHT context). This approach would ensure that the model prediction maintains robustness and efficacy even in the absence of data from certain stations or time frames, thereby enhancing the overall reliability of the reconstruction process.

\section{Conclusion} \label{sec:conclusion}

Specially constructed interferometric quantities composed of combinations of measured visibilities, called closure invariants, are immune to station-based corruptions. Although such closure quantities capture less information than traditional visibilities, they have been employed in radio astronomy for decades \citep[e.g.][]{Jennison_1958, Twiss_1960}, including in very-long baseline interferometry imaging \citep{EHT_2019_Imaging}. 

Recently, a general unified formalism of interferometric closure invariants was  developed for both co-polar and polarimetric measurements \citep{Thyagarajan_2022_CI, Samuel_2022}. The complete and independent set of co-polar closure invariants was shown to capture sufficient information to predict morphological class in simple models using trained machine learning classifiers \citep{Thyagarajan_2024_Lucas}. Subsequently, with an independently confirmed class, closure invariants could be used to parametrically reconstruct images. In this work, we show for the first time that the complete set of co-polar closure invariants in the \citet{Thyagarajan_2022_CI} formalism can be used for general-purpose direct image reconstruction of astrophysically relevant source morphologies under the constraints of sparse $uv$-coverage from very-long baseline interferometry. Below, we summarise the main results:

\begin{itemize}
    \item In this work, we train a purpose-built machine learning architecture shown in Figure \ref{fig:ML-architecture} composed of a convolutional autoencoder with a transformer encoder attachment on a training dataset composed of Gaussians, disks, rings, ellipses, $m$-rings, doubles, and CIFAR-10 with rotation, rescaling, and shearing augmentation during training. Our deep learning image reconstruction pipeline is designed to receive a set of closure invariants as input, corresponding to synthetic observations of a source at the location of M87* using a subset of the EHT VLBI array integrated over 24 hours (See Figure \ref{fig:uv-sampling} for $uv$-sampling). We call this approach to image reconstruction \direct\ for Deep learning Image Reconstruction with Closure Terms.
    \item Once trained, we test the \direct\ model's response to noise and untrained morphologies. By using only closure invariants for reconstruction, the model output is insensitive to arbitrarily large gain errors (Figure \ref{fig:gain-noise}). The reconstruction fidelity is also robust against thermal noise on the visibilities, derived from the SEFD of each station. As measured by the normalised cross-correlation score, the median reconstruction achieves 0.9 NXCORR fidelity even for untrained source morphologies (Figure \ref{fig:mixed-combinations-fidelity}). 
    \item We compare our reconstructions against state-of-the-art deconvolution and regularised maximum-likelihood forward modelling methods (standard \texttt{CLEAN}, \texttt{eht-imaging}, and \texttt{DoG-HiT}). \direct\ achieves comparable performance for both trained (Figure \ref{fig:recon-comparison}) and untrained (Figure \ref{fig:recon-comparison_untrained}) sample images. Often, even the 5th percentile reconstruction, which we use as a proxy for suboptimal performance in our model, is a reliable reconstruction with an NXCORR fidelity $>0.8$. In most cases when the source morphology is similar enough to those in its training dataset, our trained network produces sharper images with fewer artifacts compared to \texttt{CLEAN} and regularised maximum-likelihood reconstructions. 
\end{itemize}

This work has shown that by training a specially designed deep learning network, we can leverage closure invariants to directly reconstruct images from the sparse $uv$-coverage characteristic of a very-long baseline interferometric array. The results of the trained \direct\ model are insensitive to station-based corruptions, and comparable to other deconvolution and forward-modelling imaging algorithms. This work is independent and offers an additional constraint on source morphology detected by radio interferometry, ultimately improving the accuracy and reliability of sparse VLBI imaging results. In future work, \direct\ can be extended to include polarimetric imaging and increased data volume would further refine the model's reconstruction quality for diverse source morphologies.

\section*{Acknowledgements}
Inputs from Timothy Galvin, Nikhel Gupta, Li Wang, Dongjin Kim, and Mahalakshmi Sabanayagam are gratefully acknowledged. We further acknowledge Michael Janssen for providing the FITS model image of Centaurus A used in this study for validation. We also thank the anonymous referee for their constructive and insightful comments which have improved this work.

Software packages used in this study include \texttt{Numpy} \citep{Numpy_2011}, \texttt{Scipy} \citep{Scipy_2020}, \texttt{eht-imaging} \citep{Chael_2018_ehtim}, \texttt{DoG-HiT} \citep{Mueller_2022_doghit}, \texttt{PyTorch} \citep{PyTorch}, and \texttt{Matplotlib} \citep{Matplotlib_2007}.

\section*{Data Availability}
The data underlying this article will be shared on reasonable request to the corresponding author.



\bibliographystyle{mnras}
\bibliography{bibliography} 

\begin{thebibliography}{}
\makeatletter
\relax
\def\mn@urlcharsother{\let\do\@makeother \do\$\do\&\do\#\do\^\do\_\do\%\do\~}
\def\mn@doi{\begingroup\mn@urlcharsother \@ifnextchar [ {\mn@doi@} {\mn@doi@[]}}
\def\mn@doi@[#1]#2{\def\@tempa{#1}\ifx\@tempa\@empty \href {http://dx.doi.org/#2} {doi:#2}\else \href {http://dx.doi.org/#2} {#1}\fi \endgroup}
\def\mn@eprint#1#2{\mn@eprint@#1:#2::\@nil}
\def\mn@eprint@arXiv#1{\href {http://arxiv.org/abs/#1} {{\tt arXiv:#1}}}
\def\mn@eprint@dblp#1{\href {http://dblp.uni-trier.de/rec/bibtex/#1.xml} {dblp:#1}}
\def\mn@eprint@#1:#2:#3:#4\@nil{\def\@tempa {#1}\def\@tempb {#2}\def\@tempc {#3}\ifx \@tempc \@empty \let \@tempc \@tempb \let \@tempb \@tempa \fi \ifx \@tempb \@empty \def\@tempb {arXiv}\fi \@ifundefined {mn@eprint@\@tempb}{\@tempb:\@tempc}{\expandafter \expandafter \csname mn@eprint@\@tempb\endcsname \expandafter{\@tempc}}}

\bibitem[\protect\citeauthoryear{{Akiyama} et~al.,}{{Akiyama} et~al.}{2017a}]{Akiyama_2017_polarimetric}
{Akiyama} K.,  et~al., 2017a, \mn@doi [\aj] {10.3847/1538-3881/aa6302}, \href {https://ui.adsabs.harvard.edu/abs/2017AJ....153..159A} {153, 159}

\bibitem[\protect\citeauthoryear{{Akiyama} et~al.,}{{Akiyama} et~al.}{2017b}]{Akiyama_2017_imaging}
{Akiyama} K.,  et~al., 2017b, \mn@doi [\apj] {10.3847/1538-4357/aa6305}, \href {https://ui.adsabs.harvard.edu/abs/2017ApJ...838....1A} {838, 1}

\bibitem[\protect\citeauthoryear{{Arras}, {Frank}, {Haim}, {Knollm{\"u}ller}, {Leike}, {Reinecke}  \& {En{\ss}lin}}{{Arras} et~al.}{2022}]{Arras_2022}
{Arras} P.,  {Frank} P.,  {Haim} P.,  {Knollm{\"u}ller} J.,  {Leike} R.,  {Reinecke} M.,   {En{\ss}lin} T.,  2022, \mn@doi [Nature Astronomy] {10.1038/s41550-021-01548-0}, \href {https://ui.adsabs.harvard.edu/abs/2022NatAs...6..259A} {6, 259}

\bibitem[\protect\citeauthoryear{{Baars}, {Martin}, {Mangum}, {McMullin}  \& {Peters}}{{Baars} et~al.}{1999}]{Baars_1999_SMT}
{Baars} J. W.~M.,  {Martin} R.~N.,  {Mangum} J.~G.,  {McMullin} J.~P.,   {Peters} W.~L.,  1999, \mn@doi [\pasp] {10.1086/316365}, \href {https://ui.adsabs.harvard.edu/abs/1999PASP..111..627B} {111, 627}

\bibitem[\protect\citeauthoryear{{Baron}, {Monnier}  \& {Kloppenborg}}{{Baron} et~al.}{2010}]{Baron_2010SPIE}
{Baron} F.,  {Monnier} J.~D.,   {Kloppenborg} B.,  2010, in {Danchi} W.~C.,  {Delplancke} F.,   {Rajagopal} J.~K.,  eds,  Society of Photo-Optical Instrumentation Engineers (SPIE) Conference Series Vol. 7734, Optical and Infrared Interferometry II. p. 77342I, \mn@doi{10.1117/12.857364}

\bibitem[\protect\citeauthoryear{{Blackburn}, {Pesce}, {Johnson}, {Wielgus}, {Chael}, {Christian}  \& {Doeleman}}{{Blackburn} et~al.}{2020}]{Blackburn_2020}
{Blackburn} L.,  {Pesce} D.~W.,  {Johnson} M.~D.,  {Wielgus} M.,  {Chael} A.~A.,  {Christian} P.,   {Doeleman} S.~S.,  2020, \mn@doi [\apj] {10.3847/1538-4357/ab8469}, \href {https://ui.adsabs.harvard.edu/abs/2020ApJ...894...31B} {894, 31}

\bibitem[\protect\citeauthoryear{{Broderick} \& {Pesce}}{{Broderick} \& {Pesce}}{2020}]{Broderick_2020}
{Broderick} A.~E.,  {Pesce} D.~W.,  2020, \mn@doi [\apj] {10.3847/1538-4357/abbd9d}, \href {https://ui.adsabs.harvard.edu/abs/2020ApJ...904..126B} {904, 126}

\bibitem[\protect\citeauthoryear{{Broderick}, {Fish}, {Doeleman}  \& {Loeb}}{{Broderick} et~al.}{2011}]{Broderick_2011}
{Broderick} A.~E.,  {Fish} V.~L.,  {Doeleman} S.~S.,   {Loeb} A.,  2011, \mn@doi [\apj] {10.1088/0004-637X/735/2/110}, \href {https://ui.adsabs.harvard.edu/abs/2011ApJ...735..110B} {735, 110}

\bibitem[\protect\citeauthoryear{{Broderick}, {Pesce}, {Tiede}, {Pu}  \& {Gold}}{{Broderick} et~al.}{2020}]{Broderick_2020_Themis}
{Broderick} A.~E.,  {Pesce} D.~W.,  {Tiede} P.,  {Pu} H.-Y.,   {Gold} R.,  2020, \mn@doi [\apj] {10.3847/1538-4357/ab9c1f}, \href {https://ui.adsabs.harvard.edu/abs/2020ApJ...898....9B} {898, 9}

\bibitem[\protect\citeauthoryear{{Broderick} et~al.,}{{Broderick} et~al.}{2022}]{Broderick_2022}
{Broderick} A.~E.,  et~al., 2022, \mn@doi [\apj] {10.3847/1538-4357/ac7c1d}, \href {https://ui.adsabs.harvard.edu/abs/2022ApJ...935...61B} {935, 61}

\bibitem[\protect\citeauthoryear{{Carilli} \& {Thyagarajan}}{{Carilli} \& {Thyagarajan}}{2022}]{Carilli_2022}
{Carilli} C.~L.,  {Thyagarajan} N.,  2022, \mn@doi [\apj] {10.3847/1538-4357/ac3cba}, \href {https://ui.adsabs.harvard.edu/abs/2022ApJ...924..125C} {924, 125}

\bibitem[\protect\citeauthoryear{{Carlstrom} et~al.,}{{Carlstrom} et~al.}{2011}]{Carlstrom_2011_SPT}
{Carlstrom} J.~E.,  et~al., 2011, \mn@doi [\pasp] {10.1086/659879}, \href {https://ui.adsabs.harvard.edu/abs/2011PASP..123..568C} {123, 568}

\bibitem[\protect\citeauthoryear{{Chael}, {Johnson}, {Narayan}, {Doeleman}, {Wardle}  \& {Bouman}}{{Chael} et~al.}{2016}]{Chael_2016}
{Chael} A.~A.,  {Johnson} M.~D.,  {Narayan} R.,  {Doeleman} S.~S.,  {Wardle} J. F.~C.,   {Bouman} K.~L.,  2016, \mn@doi [\apj] {10.3847/0004-637X/829/1/11}, \href {https://ui.adsabs.harvard.edu/abs/2016ApJ...829...11C} {829, 11}

\bibitem[\protect\citeauthoryear{{Chael}, {Johnson}, {Bouman}, {Blackburn}, {Akiyama}  \& {Narayan}}{{Chael} et~al.}{2018}]{Chael_2018_ehtim}
{Chael} A.~A.,  {Johnson} M.~D.,  {Bouman} K.~L.,  {Blackburn} L.~L.,  {Akiyama} K.,   {Narayan} R.,  2018, \mn@doi [\apj] {10.3847/1538-4357/aab6a8}, \href {https://ui.adsabs.harvard.edu/abs/2018ApJ...857...23C} {857, 23}

\bibitem[\protect\citeauthoryear{Cornwell}{Cornwell}{2008}]{Cornwell_2008_CLEAN}
Cornwell T.~J.,  2008, \mn@doi [IEEE Journal of Selected Topics in Signal Processing] {10.1109/JSTSP.2008.2006388}, 2, 793

\bibitem[\protect\citeauthoryear{{Cornwell} \& {Evans}}{{Cornwell} \& {Evans}}{1985}]{Cornwell_1985}
{Cornwell} T.~J.,  {Evans} K.~F.,  1985, \aap, \href {https://ui.adsabs.harvard.edu/abs/1985A&A...143...77C} {143, 77}

\bibitem[\protect\citeauthoryear{{Cornwell} \& {Fomalont}}{{Cornwell} \& {Fomalont}}{1999}]{Cornwell_1999}
{Cornwell} T.,  {Fomalont} E.~B.,  1999, in {Taylor} G.~B.,  {Carilli} C.~L.,   {Perley} R.~A.,  eds,  Astronomical Society of the Pacific Conference Series Vol. 180, Synthesis Imaging in Radio Astronomy II. p.~187

\bibitem[\protect\citeauthoryear{{Doeleman} et~al.,}{{Doeleman} et~al.}{2009}]{Doeleman_2009}
{Doeleman} S.,  et~al., 2009, in astro2010: The Astronomy and Astrophysics Decadal Survey. p.~68 (\mn@eprint {arXiv} {0906.3899}), \mn@doi{10.48550/arXiv.0906.3899}

\bibitem[\protect\citeauthoryear{{Event Horizon Telescope Collaboration} et~al.,}{{Event Horizon Telescope Collaboration} et~al.}{2019a}]{EHT_2019_Array}
{Event Horizon Telescope Collaboration} et~al., 2019a, \mn@doi [\apjl] {10.3847/2041-8213/ab0c96}, \href {https://ui.adsabs.harvard.edu/abs/2019ApJ...875L...2E} {875, L2}

\bibitem[\protect\citeauthoryear{{Event Horizon Telescope Collaboration} et~al.,}{{Event Horizon Telescope Collaboration} et~al.}{2019b}]{EHT_2019_Data}
{Event Horizon Telescope Collaboration} et~al., 2019b, \mn@doi [\apjl] {10.3847/2041-8213/ab0c57}, \href {https://ui.adsabs.harvard.edu/abs/2019ApJ...875L...3E} {875, L3}

\bibitem[\protect\citeauthoryear{{Event Horizon Telescope Collaboration} et~al.,}{{Event Horizon Telescope Collaboration} et~al.}{2019c}]{EHT_2019_Imaging}
{Event Horizon Telescope Collaboration} et~al., 2019c, \mn@doi [\apjl] {10.3847/2041-8213/ab0e85}, \href {https://ui.adsabs.harvard.edu/abs/2019ApJ...875L...4E} {875, L4}

\bibitem[\protect\citeauthoryear{{Event Horizon Telescope Collaboration} et~al.,}{{Event Horizon Telescope Collaboration} et~al.}{2019d}]{EHT_2019_ShadowMass}
{Event Horizon Telescope Collaboration} et~al., 2019d, \mn@doi [\apjl] {10.3847/2041-8213/ab1141}, \href {https://ui.adsabs.harvard.edu/abs/2019ApJ...875L...6E} {875, L6}

\bibitem[\protect\citeauthoryear{{Event Horizon Telescope Collaboration} et~al.,}{{Event Horizon Telescope Collaboration} et~al.}{2022}]{EHT_2022_SgrAImaging}
{Event Horizon Telescope Collaboration} et~al., 2022, \mn@doi [\apjl] {10.3847/2041-8213/ac6429}, \href {https://ui.adsabs.harvard.edu/abs/2022ApJ...930L..14E} {930, L14}

\bibitem[\protect\citeauthoryear{{Feng}, {Bouman}  \& {Freeman}}{{Feng} et~al.}{2024}]{Feng_2024}
{Feng} B.~T.,  {Bouman} K.~L.,   {Freeman} W.~T.,  2024, \mn@doi [arXiv e-prints] {10.48550/arXiv.2406.02785}, \href {https://ui.adsabs.harvard.edu/abs/2024arXiv240602785F} {p. arXiv:2406.02785}

\bibitem[\protect\citeauthoryear{{Frieden}}{{Frieden}}{1972}]{Frieden_1972}
{Frieden} B.~R.,  1972, Journal of the Optical Society of America (1917-1983), \href {https://ui.adsabs.harvard.edu/abs/1972JOSA...62..511F} {62, 511}

\bibitem[\protect\citeauthoryear{{Garsden} et~al.,}{{Garsden} et~al.}{2015}]{Garsden_2015}
{Garsden} H.,  et~al., 2015, \mn@doi [\aap] {10.1051/0004-6361/201424504}, \href {https://ui.adsabs.harvard.edu/abs/2015A&A...575A..90G} {575, A90}

\bibitem[\protect\citeauthoryear{Gneiting, Balabdaoui  \& Raftery}{Gneiting et~al.}{2007}]{Gneiting_2007}
Gneiting T.,  Balabdaoui F.,   Raftery A.~E.,  2007, \mn@doi [Journal of the Royal Statistical Society: Series B (Statistical Methodology)] {https://doi.org/10.1111/j.1467-9868.2007.00587.x}, 69, 243

\bibitem[\protect\citeauthoryear{{Goddi} et~al.,}{{Goddi} et~al.}{2019}]{Goddi_2019_ALMA}
{Goddi} C.,  et~al., 2019, \mn@doi [\pasp] {10.1088/1538-3873/ab136a}, \href {https://ui.adsabs.harvard.edu/abs/2019PASP..131g5003G} {131, 075003}

\bibitem[\protect\citeauthoryear{{Greve} et~al.,}{{Greve} et~al.}{1995}]{Greve_1995_PV}
{Greve} A.,  et~al., 1995, \aap, \href {https://ui.adsabs.harvard.edu/abs/1995A&A...299L..33G} {299, L33}

\bibitem[\protect\citeauthoryear{{Gull} \& {Daniell}}{{Gull} \& {Daniell}}{1978}]{Gull_1978}
{Gull} S.~F.,  {Daniell} G.~J.,  1978, \mn@doi [\nat] {10.1038/272686a0}, \href {https://ui.adsabs.harvard.edu/abs/1978Natur.272..686G} {272, 686}

\bibitem[\protect\citeauthoryear{{G{\"u}sten}, {Nyman}, {Schilke}, {Menten}, {Cesarsky}  \& {Booth}}{{G{\"u}sten} et~al.}{2006}]{Gusten_2006_APEX}
{G{\"u}sten} R.,  {Nyman} L.~{\r{A}}.,  {Schilke} P.,  {Menten} K.,  {Cesarsky} C.,   {Booth} R.,  2006, \mn@doi [\aap] {10.1051/0004-6361:20065420}, \href {https://ui.adsabs.harvard.edu/abs/2006A&A...454L..13G} {454, L13}

\bibitem[\protect\citeauthoryear{{Hersbach}}{{Hersbach}}{2000}]{Hersbach_2000_CRPS}
{Hersbach} H.,  2000, \mn@doi [Weather and Forecasting] {10.1175/1520-0434(2000)015<0559:DOTCRP>2.0.CO;2}, \href {https://ui.adsabs.harvard.edu/abs/2000WtFor..15..559H} {15, 559}

\bibitem[\protect\citeauthoryear{{Ho}, {Moran}  \& {Lo}}{{Ho} et~al.}{2004}]{Ho_2004_SMA}
{Ho} P. T.~P.,  {Moran} J.~M.,   {Lo} K.~Y.,  2004, \mn@doi [\apjl] {10.1086/423245}, \href {https://ui.adsabs.harvard.edu/abs/2004ApJ...616L...1H} {616, L1}

\bibitem[\protect\citeauthoryear{{H{\"o}gbom}}{{H{\"o}gbom}}{1974}]{Hogbom_1974_CLEAN}
{H{\"o}gbom} J.~A.,  1974, \aaps, \href {https://ui.adsabs.harvard.edu/abs/1974A&AS...15..417H} {15, 417}

\bibitem[\protect\citeauthoryear{{Honma}, {Akiyama}, {Uemura}  \& {Ikeda}}{{Honma} et~al.}{2014}]{Honma_2014}
{Honma} M.,  {Akiyama} K.,  {Uemura} M.,   {Ikeda} S.,  2014, \mn@doi [\pasj] {10.1093/pasj/psu070}, \href {https://ui.adsabs.harvard.edu/abs/2014PASJ...66...95H} {66, 95}

\bibitem[\protect\citeauthoryear{{Hughes} et~al.,}{{Hughes} et~al.}{2010}]{Hughes_2010_LMT}
{Hughes} D.~H.,  et~al., 2010, in {Stepp} L.~M.,  {Gilmozzi} R.,   {Hall} H.~J.,  eds,  Society of Photo-Optical Instrumentation Engineers (SPIE) Conference Series Vol. 7733, Ground-based and Airborne Telescopes III. p. 773312, \mn@doi{10.1117/12.857974}

\bibitem[\protect\citeauthoryear{{Hunter}}{{Hunter}}{2007}]{Matplotlib_2007}
{Hunter} J.~D.,  2007, \mn@doi [Computing in Science and Engineering] {10.1109/MCSE.2007.55}, \href {https://ui.adsabs.harvard.edu/abs/2007CSE.....9...90H} {9, 90}

\bibitem[\protect\citeauthoryear{{Ikeda}, {Tazaki}, {Akiyama}, {Hada}  \& {Honma}}{{Ikeda} et~al.}{2016}]{Ikeda_2016}
{Ikeda} S.,  {Tazaki} F.,  {Akiyama} K.,  {Hada} K.,   {Honma} M.,  2016, \mn@doi [\pasj] {10.1093/pasj/psw042}, \href {https://ui.adsabs.harvard.edu/abs/2016PASJ...68...45I} {68, 45}

\bibitem[\protect\citeauthoryear{{Janssen} et~al.,}{{Janssen} et~al.}{2021}]{Janssen_2021_CenA}
{Janssen} M.,  et~al., 2021, \mn@doi [Nature Astronomy] {10.1038/s41550-021-01417-w}, \href {https://ui.adsabs.harvard.edu/abs/2021NatAs...5.1017J} {5, 1017}

\bibitem[\protect\citeauthoryear{{Jennison}}{{Jennison}}{1958}]{Jennison_1958}
{Jennison} R.~C.,  1958, \mn@doi [\mnras] {10.1093/mnras/118.3.276}, \href {https://ui.adsabs.harvard.edu/abs/1958MNRAS.118..276J} {118, 276}

\bibitem[\protect\citeauthoryear{{Kim} et~al.,}{{Kim} et~al.}{2018}]{Kim_2018_SPT}
{Kim} J.,  et~al., 2018, in {Zmuidzinas} J.,  {Gao} J.-R.,  eds,  Society of Photo-Optical Instrumentation Engineers (SPIE) Conference Series Vol. 10708, Millimeter, Submillimeter, and Far-Infrared Detectors and Instrumentation for Astronomy IX. p. 107082S (\mn@eprint {arXiv} {1805.09346}), \mn@doi{10.1117/12.2301005}

\bibitem[\protect\citeauthoryear{{Kim} et~al.,}{{Kim} et~al.}{2020}]{Kim_2020_3C279}
{Kim} J.-Y.,  et~al., 2020, \mn@doi [\aap] {10.1051/0004-6361/202037493}, \href {https://ui.adsabs.harvard.edu/abs/2020A&A...640A..69K} {640, A69}

\bibitem[\protect\citeauthoryear{{Kingma} \& {Ba}}{{Kingma} \& {Ba}}{2014}]{Kingma_2014}
{Kingma} D.~P.,  {Ba} J.,  2014, \mn@doi [arXiv e-prints] {10.48550/arXiv.1412.6980}, \href {https://ui.adsabs.harvard.edu/abs/2014arXiv1412.6980K} {p. arXiv:1412.6980}

\bibitem[\protect\citeauthoryear{Kramer}{Kramer}{1992}]{KRAMER1992313}
Kramer M.,  1992, \mn@doi [Computers & Chemical Engineering] {https://doi.org/10.1016/0098-1354(92)80051-A}, 16, 313

\bibitem[\protect\citeauthoryear{Krizhevsky}{Krizhevsky}{2009}]{Krizhevsky09_CIFAR10}
Krizhevsky A.,  2009, Technical report, Learning multiple layers of features from tiny images.
University of Toronto, Toronto, Ontario

\bibitem[\protect\citeauthoryear{{Kuramochi}, {Akiyama}, {Ikeda}, {Tazaki}, {Fish}, {Pu}, {Asada}  \& {Honma}}{{Kuramochi} et~al.}{2018}]{Kuramochi_2018}
{Kuramochi} K.,  {Akiyama} K.,  {Ikeda} S.,  {Tazaki} F.,  {Fish} V.~L.,  {Pu} H.-Y.,  {Asada} K.,   {Honma} M.,  2018, \mn@doi [\apj] {10.3847/1538-4357/aab6b5}, \href {https://ui.adsabs.harvard.edu/abs/2018ApJ...858...56K} {858, 56}

\bibitem[\protect\citeauthoryear{Lai}{Lai}{2024}]{DIReCT_Zenodo}
Lai S.,  2024, samlaihei/DIReCT: v1.0.0 DIReCT, \mn@doi{10.5281/zenodo.14032844}, \url {https://doi.org/10.5281/zenodo.14032844}

\bibitem[\protect\citeauthoryear{{Li}, {Cornwell}  \& {de Hoog}}{{Li} et~al.}{2011}]{Li_2011_CompSens}
{Li} F.,  {Cornwell} T.~J.,   {de Hoog} F.,  2011, \mn@doi [\aap] {10.1051/0004-6361/201015045}, \href {https://ui.adsabs.harvard.edu/abs/2011A&A...528A..31L} {528, A31}

\bibitem[\protect\citeauthoryear{{Lister}, {Aller}, {Aller}, {Hodge}, {Homan}, {Kovalev}, {Pushkarev}  \& {Savolainen}}{{Lister} et~al.}{2018}]{Lister_2018}
{Lister} M.~L.,  {Aller} M.~F.,  {Aller} H.~D.,  {Hodge} M.~A.,  {Homan} D.~C.,  {Kovalev} Y.~Y.,  {Pushkarev} A.~B.,   {Savolainen} T.,  2018, \mn@doi [\apjs] {10.3847/1538-4365/aa9c44}, \href {https://ui.adsabs.harvard.edu/abs/2018ApJS..234...12L} {234, 12}

\bibitem[\protect\citeauthoryear{Liu, Luo, Wang  \& Tang}{Liu et~al.}{2015}]{Liu_2015_CelebA}
Liu Z.,  Luo P.,  Wang X.,   Tang X.,  2015, in 2015 IEEE International Conference on Computer Vision (ICCV). pp 3730--3738, \mn@doi{10.1109/ICCV.2015.425}

\bibitem[\protect\citeauthoryear{{Lockhart} \& {Gralla}}{{Lockhart} \& {Gralla}}{2022}]{Lockhart_2022}
{Lockhart} W.,  {Gralla} S.~E.,  2022, \mn@doi [\mnras] {10.1093/mnras/stab3204}, \href {https://ui.adsabs.harvard.edu/abs/2022MNRAS.509.3643L} {509, 3643}

\bibitem[\protect\citeauthoryear{{Marsh} \& {Richardson}}{{Marsh} \& {Richardson}}{1987}]{Marsh_1987}
{Marsh} K.~A.,  {Richardson} J.~M.,  1987, \aap, \href {https://ui.adsabs.harvard.edu/abs/1987A&A...182..174M} {182, 174}

\bibitem[\protect\citeauthoryear{{Mertens} \& {Lobanov}}{{Mertens} \& {Lobanov}}{2015}]{Mertens_2015}
{Mertens} F.,  {Lobanov} A.,  2015, \mn@doi [\aap] {10.1051/0004-6361/201424566}, \href {https://ui.adsabs.harvard.edu/abs/2015A&A...574A..67M} {574, A67}

\bibitem[\protect\citeauthoryear{{M{\"u}ller}}{{M{\"u}ller}}{2024}]{Muller_2024_closureTraces}
{M{\"u}ller} H.,  2024, \mn@doi [arXiv e-prints] {10.48550/arXiv.2407.20190}, \href {https://ui.adsabs.harvard.edu/abs/2024arXiv240720190M} {p. arXiv:2407.20190}

\bibitem[\protect\citeauthoryear{{M{\"u}ller} \& {Lobanov}}{{M{\"u}ller} \& {Lobanov}}{2022}]{Mueller_2022_doghit}
{M{\"u}ller} H.,  {Lobanov} A.~P.,  2022, \mn@doi [\aap] {10.1051/0004-6361/202243244}, \href {https://ui.adsabs.harvard.edu/abs/2022A&A...666A.137M} {666, A137}

\bibitem[\protect\citeauthoryear{{M{\"u}ller}, {Mus}  \& {Lobanov}}{{M{\"u}ller} et~al.}{2023}]{Muller_2023_Multiobj}
{M{\"u}ller} H.,  {Mus} A.,   {Lobanov} A.,  2023, \mn@doi [\aap] {10.1051/0004-6361/202346207}, \href {https://ui.adsabs.harvard.edu/abs/2023A&A...675A..60M} {675, A60}

\bibitem[\protect\citeauthoryear{{Narayan} \& {Nityananda}}{{Narayan} \& {Nityananda}}{1986}]{Narayan_1986}
{Narayan} R.,  {Nityananda} R.,  1986, \mn@doi [\araa] {10.1146/annurev.aa.24.090186.001015}, \href {https://ui.adsabs.harvard.edu/abs/1986ARA&A..24..127N} {24, 127}

\bibitem[\protect\citeauthoryear{{Offringa} et~al.,}{{Offringa} et~al.}{2014}]{Offringa_2014_WSCLEAN}
{Offringa} A.~R.,  et~al., 2014, \mn@doi [\mnras] {10.1093/mnras/stu1368}, \href {https://ui.adsabs.harvard.edu/abs/2014MNRAS.444..606O} {444, 606}

\bibitem[\protect\citeauthoryear{{Paszke} et~al.,}{{Paszke} et~al.}{2019}]{PyTorch}
{Paszke} A.,  et~al., 2019, \mn@doi [arXiv e-prints] {10.48550/arXiv.1912.01703}, \href {https://ui.adsabs.harvard.edu/abs/2019arXiv191201703P} {p. arXiv:1912.01703}

\bibitem[\protect\citeauthoryear{{Polsterer}}{{Polsterer}}{2017}]{Polsterer_2017}
{Polsterer} K.~L.,  2017, in {Brescia} M.,  {Djorgovski} S.~G.,  {Feigelson} E.~D.,  {Longo} G.,   {Cavuoti} S.,  eds,  IAU Symposium Vol. 325, Astroinformatics. pp 156--165, \mn@doi{10.1017/S1743921316013089}

\bibitem[\protect\citeauthoryear{{Polsterer} \& {Gieseke}}{{Polsterer} \& {Gieseke}}{2019}]{Polsterer_2019}
{Polsterer} K.~L.,  {Gieseke} F.,  2019, in {Molinaro} M.,  {Shortridge} K.,   {Pasian} F.,  eds,  Astronomical Society of the Pacific Conference Series Vol. 521, Astronomical Data Analysis Software and Systems XXVI. p.~240

\bibitem[\protect\citeauthoryear{{Pratley}, {McEwen}, {d'Avezac}, {Carrillo}, {Onose}  \& {Wiaux}}{{Pratley} et~al.}{2018}]{Pratley_2018}
{Pratley} L.,  {McEwen} J.~D.,  {d'Avezac} M.,  {Carrillo} R.~E.,  {Onose} A.,   {Wiaux} Y.,  2018, \mn@doi [\mnras] {10.1093/mnras/stx2237}, \href {https://ui.adsabs.harvard.edu/abs/2018MNRAS.473.1038P} {473, 1038}

\bibitem[\protect\citeauthoryear{{Readhead} \& {Wilkinson}}{{Readhead} \& {Wilkinson}}{1978}]{Readhead_1978}
{Readhead} A.~C.~S.,  {Wilkinson} P.~N.,  1978, \mn@doi [\apj] {10.1086/156232}, \href {https://ui.adsabs.harvard.edu/abs/1978ApJ...223...25R} {223, 25}

\bibitem[\protect\citeauthoryear{{Roelofs} et~al.,}{{Roelofs} et~al.}{2023}]{Roelofs_2023_mring}
{Roelofs} F.,  et~al., 2023, \mn@doi [\apjl] {10.3847/2041-8213/acff6f}, \href {https://ui.adsabs.harvard.edu/abs/2023ApJ...957L..21R} {957, L21}

\bibitem[\protect\citeauthoryear{{Rogers} et~al.,}{{Rogers} et~al.}{1974}]{Rogers_1974}
{Rogers} A.~E.~E.,  et~al., 1974, \mn@doi [\apj] {10.1086/153162}, \href {https://ui.adsabs.harvard.edu/abs/1974ApJ...193..293R} {193, 293}

\bibitem[\protect\citeauthoryear{Rudin, Osher  \& Fatemi}{Rudin et~al.}{1992}]{RUDIN1992259}
Rudin L.~I.,  Osher S.,   Fatemi E.,  1992, \mn@doi [Physica D: Nonlinear Phenomena] {https://doi.org/10.1016/0167-2789(92)90242-F}, 60, 259

\bibitem[\protect\citeauthoryear{{Samuel}, {Nityananda}  \& {Thyagarajan}}{{Samuel} et~al.}{2022}]{Samuel_2022}
{Samuel} J.,  {Nityananda} R.,   {Thyagarajan} N.,  2022, \mn@doi [\prl] {10.1103/PhysRevLett.128.091101}, \href {https://ui.adsabs.harvard.edu/abs/2022PhRvL.128i1101S} {128, 091101}

\bibitem[\protect\citeauthoryear{{Schwarz}}{{Schwarz}}{1978}]{Schwarz_1978}
{Schwarz} U.~J.,  1978, \aap, \href {https://ui.adsabs.harvard.edu/abs/1978A&A....65..345S} {65, 345}

\bibitem[\protect\citeauthoryear{{Shepherd}}{{Shepherd}}{2011}]{Shepherd_2011_DIFMAP}
{Shepherd} M.,  2011, {Difmap: Synthesis Imaging of Visibility Data}, Astrophysics Source Code Library, record ascl:1103.001

\bibitem[\protect\citeauthoryear{{Sun} \& {Bouman}}{{Sun} \& {Bouman}}{2020}]{Sun_2020}
{Sun} H.,  {Bouman} K.~L.,  2020, \mn@doi [arXiv e-prints] {10.48550/arXiv.2010.14462}, \href {https://ui.adsabs.harvard.edu/abs/2020arXiv201014462S} {p. arXiv:2010.14462}

\bibitem[\protect\citeauthoryear{{Sun}, {Bouman}, {Tiede}, {Wang}, {Blunt}  \& {Mawet}}{{Sun} et~al.}{2022}]{Sun_2022}
{Sun} H.,  {Bouman} K.~L.,  {Tiede} P.,  {Wang} J.~J.,  {Blunt} S.,   {Mawet} D.,  2022, \mn@doi [\apj] {10.3847/1538-4357/ac6be9}, \href {https://ui.adsabs.harvard.edu/abs/2022ApJ...932...99S} {932, 99}

\bibitem[\protect\citeauthoryear{{Taylor}, {Carilli}  \& {Perley}}{{Taylor} et~al.}{1999}]{Taylor_SIRAII_1999}
{Taylor} G.~B.,  {Carilli} C.~L.,   {Perley} R.~A.,  eds, 1999, {Synthesis Imaging in Radio Astronomy II}  Astronomical Society of the Pacific Conference Series Vol. 180

\bibitem[\protect\citeauthoryear{{Thompson}, {Moran}  \& {Swenson}}{{Thompson} et~al.}{2017}]{TMS}
{Thompson} A.~R.,  {Moran} J.~M.,   {Swenson} George~W. J.,  2017, {Interferometry and Synthesis in Radio Astronomy, 3rd Edition}.
Astronomy \& Astrophysics Library, \mn@doi{10.1007/978-3-319-44431-4}

\bibitem[\protect\citeauthoryear{{Thyagarajan}, {Nityananda}  \& {Samuel}}{{Thyagarajan} et~al.}{2022}]{Thyagarajan_2022_CI}
{Thyagarajan} N.,  {Nityananda} R.,   {Samuel} J.,  2022, \mn@doi [\prd] {10.1103/PhysRevD.105.043019}, \href {https://ui.adsabs.harvard.edu/abs/2022PhRvD.105d3019T} {105, 043019}

\bibitem[\protect\citeauthoryear{{Thyagarajan}, {Hoefs}  \& {Wong}}{{Thyagarajan} et~al.}{2024}]{Thyagarajan_2024_Lucas}
{Thyagarajan} N.,  {Hoefs} L.,   {Wong} O.~I.,  2024, \mn@doi [RAS Techniques and Instruments] {10.1093/rasti/rzae031}, \href {https://ui.adsabs.harvard.edu/abs/2024RASTI...3..437T} {3, 437}

\bibitem[\protect\citeauthoryear{{Twiss}, {Carter}  \& {Little}}{{Twiss} et~al.}{1960}]{Twiss_1960}
{Twiss} R.~Q.,  {Carter} A.~W.~L.,   {Little} A.~G.,  1960, The Observatory, \href {https://ui.adsabs.harvard.edu/abs/1960Obs....80..153T} {80, 153}

\bibitem[\protect\citeauthoryear{{Vaswani}, {Shazeer}, {Parmar}, {Uszkoreit}, {Jones}, {Gomez}, {Kaiser}  \& {Polosukhin}}{{Vaswani} et~al.}{2017}]{Vaswani_2017}
{Vaswani} A.,  {Shazeer} N.,  {Parmar} N.,  {Uszkoreit} J.,  {Jones} L.,  {Gomez} A.~N.,  {Kaiser} L.,   {Polosukhin} I.,  2017, \mn@doi [arXiv e-prints] {10.48550/arXiv.1706.03762}, \href {https://ui.adsabs.harvard.edu/abs/2017arXiv170603762V} {p. arXiv:1706.03762}

\bibitem[\protect\citeauthoryear{{Virtanen} et~al.,}{{Virtanen} et~al.}{2020}]{Scipy_2020}
{Virtanen} P.,  et~al., 2020, \mn@doi [Nature Methods] {10.1038/s41592-019-0686-2}, \href {https://ui.adsabs.harvard.edu/abs/2020NatMe..17..261V} {17, 261}

\bibitem[\protect\citeauthoryear{{Wakker} \& {Schwarz}}{{Wakker} \& {Schwarz}}{1988}]{Wakker_1988}
{Wakker} B.~P.,  {Schwarz} U.~J.,  1988, \aap, \href {https://ui.adsabs.harvard.edu/abs/1988A&A...200..312W} {200, 312}

\bibitem[\protect\citeauthoryear{{Wiaux}, {Jacques}, {Puy}, {Scaife}  \& {Vandergheynst}}{{Wiaux} et~al.}{2009}]{Wiaux_2009}
{Wiaux} Y.,  {Jacques} L.,  {Puy} G.,  {Scaife} A.~M.~M.,   {Vandergheynst} P.,  2009, \mn@doi [\mnras] {10.1111/j.1365-2966.2009.14665.x}, \href {https://ui.adsabs.harvard.edu/abs/2009MNRAS.395.1733W} {395, 1733}

\bibitem[\protect\citeauthoryear{{Wielgus} et~al.,}{{Wielgus} et~al.}{2022}]{Wielgus_2022}
{Wielgus} M.,  et~al., 2022, \mn@doi [\aap] {10.1051/0004-6361/202244493}, \href {https://ui.adsabs.harvard.edu/abs/2022A&A...665L...6W} {665, L6}

\bibitem[\protect\citeauthoryear{{Wootten} \& {Thompson}}{{Wootten} \& {Thompson}}{2009}]{Wootten_2009_ALMA}
{Wootten} A.,  {Thompson} A.~R.,  2009, \mn@doi [IEEE Proceedings] {10.1109/JPROC.2009.2020572}, \href {https://ui.adsabs.harvard.edu/abs/2009IEEEP..97.1463W} {97, 1463}

\bibitem[\protect\citeauthoryear{{van der Walt}, {Colbert}  \& {Varoquaux}}{{van der Walt} et~al.}{2011}]{Numpy_2011}
{van der Walt} S.,  {Colbert} S.~C.,   {Varoquaux} G.,  2011, \mn@doi [Computing in Science and Engineering] {10.1109/MCSE.2011.37}, \href {https://ui.adsabs.harvard.edu/abs/2011CSE....13b..22V} {13, 22}

\makeatother
\end{thebibliography}






\bsp	
\label{lastpage}
\end{document}